\documentstyle[12pt]{article}
\newcommand{\beq}{\begin{equation}}
\newcommand{\eeq}{\end{equation}}

\def\sqrtwo{{\textstyle{{\sqrt{2}}}}}

\def\betag{{\textstyle {{{1}\over{g^2}}} }}
\def\half{{\textstyle{1\over2}}}
\def\quart{{\textstyle{1\over4}}}

\def\kap{{\textstyle{1\over{\kappa^2}}}}
\def\gap{{\textstyle{1\over{g^2}}}}
\def\half{{\textstyle{1\over2}}}
\def\third{{\textstyle{1\over3}}}

\def\quart{{\textstyle{1\over4}}}
\def\eigth{{\textstyle{{{1}\over{8}}}}}

\def\twothird{{\textstyle{2\over3}}}

\def\ninsix{{\textstyle{{{1}\over{3\cdot 2^6}}}}}

\def\thre{{\textstyle{1\over3}}}

\def\p1half{{\textstyle{{{p+1}\over{2}}}}}

\def\23phalf{{\textstyle{{{23-p}\over{2}}}}}

    \let\p=\pi

 \def\bd{\begin{document}} \def\ed{\end{document}}
\def\ds{\documentstyle} \let\fr=\frac \let\bl=\bigl \let\br=\bigr
\let\Br=\Bigr \let\Bl=\Bigl
\let\bm=\bibitem
\let\na=\nabla
\let\pa=\partial \let\ov=\overline
\newcommand{\be}{\begin{equation}}
\newcommand{\ee}{\end{equation}}
\def\ba{\begin{array}}
\def\ea{\end{array}}
\def\ft#1#2{{\textstyle{{\scriptstyle #1}\over {\scriptstyle #2}}}}
\def\fft#1#2{{#1 \over #2}}
\def\del{\partial}
\def\sst#1{{\scriptscriptstyle #1}}
\def\oneone{\rlap 1\mkern4mu{\rm l}}
\def\ie{{\it i.e.\ }}
\begin{document}
\thispagestyle{empty}
\begin{titlepage}

\bigskip
\hskip 3.7in{\vbox{\baselineskip12pt
}}

\bigskip\bigskip\bigskip\bigskip
\centerline{\large\bf Spacetime Reduction of Large N Flavor
Models:} \centerline{\large\bf A Fundamental Theory of Emergent
Local Geometry?}

\bigskip\bigskip
\bigskip\bigskip
\centerline{\bf Shyamoli Chaudhuri \footnote{Current Address: 1312
Oak Drive, Blacksburg, VA 24060. E-mail: shyamolic@yahoo.com}}

\medskip
\medskip
\medskip

\centerline{214 North Allegheny Street}
\centerline{Bellefonte, PA 16823, USA}

\bigskip

\date{\today}

\bigskip\bigskip
\begin{abstract}
\noindent We introduce a novel spacetime reduction procedure for
the fields of a supergravity-Yang-Mills theory in generic curved
spacetime background, and with large $N$ flavor group, to
linearized forms on an infinitesimal patch of local tangent space
at a point in the spacetime manifold. Our new prescription for
spacetime reduction preserves all of the local symmetries of the
continuum field theory Lagrangian in the resulting
zero-dimensional matrix Lagrangian, thereby obviating difficulties
encountered in previous matrix proposals for emergent spacetime in
recovering the full nonlinear symmetries of Einstein gravity. It
also obviates the challenges that must be faced by any proposal
for a fundamental theory, holographic or topological, where
gravity emerges instead as an induced interaction. We conjecture
that the zero-dimensional matrix model obtained by this
prescription for spacetime reduction of the circle-compactified
type I-I$^{\prime}$-mIIA-IIB-heterotic supergravity-Yang-Mills
theory with sixteen supercharges and large $N$ flavor group, and
inclusive of the full spectrum of Dpbrane charges, $-2$ $\le$ $p$
$\le$ $9$, offers a potentially complete framework for
nonperturbative string/M theory. We analyze the matrix Lagrangian
in detail, comparing with the results of traditional planar
reduction, and clarifying the emergence of the spacetime continuum
in the large $N$ limit of the zero dimensional matrix model. We
explain the relationship of our conjecture for a fundamental
theory of emergent local spacetime geometry to recent
investigations of the hidden symmetry algebra of M theory,
stressing insights that are to be gained from the algebraic
perspective. We conclude with a list of open questions and
directions for future work.
\end{abstract}
\end{titlepage}

\section{Introduction}

\vskip 0.1in In this paper, we present a zero-dimensional matrix
Lagrangian with sixteen supercharges, and an extended symmetry
group, which we will argue has the potential to describe a
multi-dimensional emergent local spacetime geometry in the large N
limit, and which might also give a unified framework for all of
the weakly-coupled string and field theory limits of
nonperturbative string/M theory. This new matrix Lagrangian
belongs to a much more general class of matrix models conjectured
to exist on the basis of algebraic symmetries alone in
\cite{mat1}.\footnote{In the analysis of the matrix superalgebra
at finite $N$ outlined in \cite{mat1}, the parameters for
infinitesimal supersymmetry and $SL(n,{\bf R})$ transformations
were assumed, more generally, to transform as non-singlets under
the flavor $U(N)$ group. While we know of no reason to rule out
such exotic extensions from the perspective of the matrix
Lagrangian, such an extension is not necessary for the problem at
hand. We should clarify that the matrix Lagrangian given in
\cite{mat1} was guessed at by intuition alone; the detailed checks
of the symmetry transformations necessary to confirm the closure
of such an exotic matrix algebra from first principles are quite
beyond this author's ability, although we encourage the reader to
try. An alternative route that can reproduce the form of the
matrix Lagrangian first given in \cite{mat1} is provided in this
paper, based upon the spacetime reduction of a
supergravity-Yang-Mills Lagrangian with large $N$ flavor symmetry.
Such a derivation obviates the necessity for a first principles
check of the matrix superalgebra at finite $N$, while also
illuminating the connection to continuum physics in the large $N$
limit. Notice that, for the more general class of matrix
superalgebras described in \cite{mat1}, the large $N$ limit would
have to correspond to an exotic (nonlinear) extension of the Nahm
classification of spacetime linear superalgebras in $n$ target
spacetime dimensions. I am grateful to Bernard de Witt for
pointing out this distinction. The idea of motivating the matrix
Lagrangian from a generalized spacetime reduction procedure,
combining insights from both supergravity dimensional reduction
\cite{cj,cjlp} and Eguchi-Kawai rigid Yang-Mills planar reduction
\cite{ek}, was inspired by Hermann Nicolai's comments in
\cite{nicolai}.} The matrix Lagrangian discussed in this paper is
therefore distinguished by its simplicity: the $U(N)$ symmetry is
assumed to be a flavor symmetry, commuting with the global
symmetries that assume the role of target spacetime
(supersymmetry) $\times$ (Lorentz) $\times$ (Yang-Mills) symmetry
in the continuum target-space Lagrangian obtained in the large $N$
limit.

\vskip 0.1in In what follows, we will explain the appearance of
such novel matrix Lagrangians with extended symmetry from the
perspective of supergravity hidden symmetry algebras
 \cite{cj,cjlp}, on the one hand, and from the notion of a modified
prescription for the planar reduction of large N field theories
\cite{ek} on the other. We will show that a subset of the matrix
Lagrangians introduced by us in \cite{mat1} can be derived by a
novel spacetime reduction procedure for the spacetime fields of a
higher-dimensional supergravity-Yang-Mills field theory to
linearized forms on an infinitesimal patch of local tangent space
at a single point in the spacetime manifold. The result of our new
prescription for spacetime reduction of a continuum Lagrangian
with large $N$ flavor symmetry is in each case a zero dimensional
matrix model Lagrangian, where the large N index originates in the
auxiliary large N flavor symmetry introduced in the higher
dimensional supergravity-Yang-Mills theory. We should emphasize
that the quantum dynamics of the higher dimensional classical
field theory Lagrangian with large N flavor group is of {\em no}
interest to us; the introduction of the auxiliary flavor index in
the continuum Lagrangian is simply {\em a technical device} that
enables straightforward derivation of a zero-dimensional matrix
Lagrangian manifesting certain desired extended symmetries.
Hopefully, authors other than myself will succeed in simply doing
away with the \lq\lq crutch" of {\em spacetime reduction}, writing
down the zero-dimensional matrix Lagrangian for the proposed
fundamental theory of emergent spacetime geometry from first
principles on the basis of algebraic symmetries alone, as
explained in Footnote 2. We reiterate that it is the large N
quantum dynamics of this matrix Lagrangian that is of fundamental
physical significance to nonperturbative String/M theory.

\vskip 0.1in In what follows, we will present a novel prescription
for the spacetime reduction of the continuum fields of a
supergravity-Yang-Mills theory in generic curved spacetime
background, and with large $N$ flavor group, to linearized
forms\footnote{By {\em linearized} pforms, we mean that each
spacetime field, or associated pform, is truncated at linear order
in a Taylor expansion about its value at the origin of the
infinitesimal patch of tangent space, coincident with a single
point in the spacetime manifold. The truncation at linear order is
simply a reflection of the fact that tangent space has
infinitesimal extent: higher order partial derivatives exist to
all orders, but the Taylor expansion is truncated because
quadratic and higher powers of the infinitesimal in tangent space
are dropped. For more details, see Section 3.} on an infinitesimal
patch of local tangent space at a single point in the spacetime
manifold, and which results in a new class of zero-dimensional
supermatrix models first presented by us in \cite{mat1}. Our novel
prescription combines insights from both supergravity dimensional
reduction \cite{cj,cjlp} as well as the well-known Eguchi-Kawai
planar reduction of rigid Yang-Mills theories \cite{ek}, extended
to the case of rigid supersymmetric large N Yang-Mills in the IKKT
matrix model \cite{ikkt}. We should clarify that, unlike the IKKT
matrix model which corresponds to a modification of the planar
reduced rigidly supersymmetric large $N$ Yang-Mills theory, the
class of matrix models studied in this paper, and in \cite{mat1},
arise from the spacetime reduction of a higher-dimensional
continuum Lagrangian with large $N$ flavor symmetry. In each case,
the target spacetime (supersymmetry) $\times$ (Lorentz) $\times$
(Yang-Mills gauge) symmetries commute with $U(N)$ flavor symmetry.
Notice, also, that we are proposing an entirely different origin
for continuum target spacetime symmetries from that conjectured by
Iso and Kawai \cite{isok}. Their idea was that continuum target
spacetime symmetries, such as the finite-dimensional Yang-Mills
gauge symmetry of the continuum Lagrangian, should already be
present in the matrix Lagrangian, embedded within the large $N$
gauge group. This proposal is difficult to realize in practice for
standard choices of the large $N$ gauge group as can be seen by
simple group theoretic reasoning.\footnote{I thank B.\ de Witt for
a discussion of this point.}

\vskip 0.1in Let us begin with a quick survey of some pertinent
insights from the relevant older works. It is well-known that the
toroidally-compactified eleven-dimensional supergravity, as well
as the ten-dimensional type I-I$^{\prime}$, type IIA-IIB, and the
heterotic $E_8$$\times$$E_8$ and $Spin(32)/{\rm Z}_2$, string
supergravities, exhibit extended global symmetries as a
consequence of the presence of massless scalar fields in the
dimensionally-reduced supergravity Lagrangian. In 1978, Cremmer
and Julia noticed that the dimensional reduction of a
$(D$$+$$n)$-dimensional theory containing gravity to $D$
dimensions necessarily results in the appearance of an $SL(n,{\bf
R})$ global symmetry, as viewed from the perspective of the
$D$-dimensional spacetime \cite{cj}. This symmetry is manifest in
the form of the dimensionally-reduced Lagrangian. Including an
overall scaling of the volume of the compactification manifold,
the global symmetry group of the Lagrangian takes the precise form
$GL(n, {\bf R})$$\sim$$SL(n, {\bf R})$$ \times $${\bf R}$; the
${\bf R}$ factor is, therefore, a {\em hidden} symmetry of the
dimensionally-reduced Lagrangian. Notice that only the volume
preserving subgroup, $SL(n,{\bf R})$, is relevant to the reduction
of a field theory to a single spacetime point. Recall that
eleven-dimensional supergravity is one of the field theoretic low
energy limits of M theory. In \cite{cj}, Cremmer and Julia
conjectured, with partial proof, that the dimensional reduction of
11d supergravity to a Lagrangian in $11$$-$$n$ dimensions would
result in the appearance of the hidden symmetry group $E_{n}$. For
$n$ $\ge$ $3$, this conjecture has since been verified by direct
field-theoretic duality transformations on the fields in the
dimensionally-reduced classical supergravity Lagrangian
\cite{cjlp}. Finally, note that if we continue Cremmer and Julia's
sequence of dimensional reductions of 11d supergravity to lower
dimensions to its logical endpoint, namely, to {\em zero}
spacetime dimensions, we have the prediction $E_{11}$ for the
hidden symmetry group. Thus, it is natural to expect that the
extended symmetry group of the zero-dimensional matrix model
obtained from the spacetime reduction of a theory with 32
supercharges would be $E_{11}$ \cite{w4,matn6}. This simple
observation will be developed, and refined further, in sections 4
and 5 of this paper.

\vskip 0.1in Planar reduction was first applied to the bosonic
rigid large $N$ Yang-Mills theory by Eguchi and Kawai in 1980
\cite{ek}. Dimensional reduction of a rigid $U(N)$ Yang-Mills
gauge theory to a single spacetime point gives what is known as a
reduced unitary matrix model: naively, we set to zero all
spacetime derivatives in the Yang-Mills action, retaining the
$U(N)$ trace of the square of the commutator of $N$$\times$$N$
unitary matrices. The Lagrangian gives a zero-dimensional unitary
matrix model with a quartic self-interaction.\footnote{We use the
term {\em Lagrangian} for a reduced matrix model as follows: the
Feynman path integral describing the quantum mechanics of the
matrix model is a sum over matrix configurations, weighted by an
exponentiated matrix-valued function. In analogy with field
theory, we will refer to the exponentiated function defined on the
field of unitary matrices, as the matrix {\em Lagrangian}. By
contrast, in the case of the famous c=1 matrix model, the matrix
variables were taken to be functions of an auxiliary parameter
later identified with target-space time \cite{mm}. Thus, the
exponentiated function weighting the Feynman path integral in the
case of the c=1 matrix model is the {\em action}, expressed as a
one-dimensional integral over time.} Reduced matrix models arise,
therefore, as the result of a dramatic thinning of the infinite
number of degrees of freedom of a quantum field theory upon
dimensional reduction of all spacetime fields to a single
spacetime point. Remarkably, planar reduced matrix models are
found to share many features of exactly solvable unitary matrix
models. It should be emphasized that many of the notions familiar
from continuum quantum field theory, such as renormalization,
universality classes, vacuum structure, and spontaneous symmetry
breaking, have their counterpart in the matrix models that follow
from spacetime reduction. Likewise, {\em super}matrix models are
obtained when one dimensionally reduces a rigid supersymmetric
large $N$ Yang-Mills theory to a spacetime point. Such supermatrix
models have been the basis of previous conjectures for
nonperturbative string/M theory \cite{bfss,ikkt}.

\vskip 0.1in With the discovery of Dirichlet-pbranes by Dai, Leigh
and Polchinski \cite{dlp}, and with their crucial role as
solitonic carriers of dual electric-magnetic charge in the type I
and type II string supergravities clarified by Polchinski in
\cite{dbrane}, the dimensional reduction of rigid Yang-Mills
theories has found an alternative, and rather interesting, new
interpretation. Recall that in open and closed string theories,
$n$ successive T-duality transformations on $n$ spacetime
coordinates parallel to the worldvolume of $N$ coincident D9branes
in the type IB string theory carrying 10d nonabelian Yang-Mills
gauge fields: $R_n$ $\to$ $\alpha^{\prime}/R_n$, where $n$ $\le$
$10$, converts $n$ components of the worldvolume gauge bosons to
the $n$ components of a scalar field in the $n$-dimensional
spatial bulk orthogonal to the D(9-n)brane \cite{dlp}. The vacuum
expectation values of the $n$ components of the scalar field can
be interpreted as the coordinate locations of the D(9-n)brane
soliton in an $n$-dimensional space. In open string theory, this
scalar excitation has as vertex operator $(\partial_{z}
X^{i}_{\mu} )^2$, where $\mu$$=$$1$, $\cdots$, $n$, and $i$$=$$1$,
$\cdots$, $N$. As first noted by Witten, this implied the
tantalizing fact that the \lq\lq coordinates" of space orthogonal
to the $N$ D(9-n)branes arise as the $N$ eigenvalues of $n$
noncommuting, $N$$\times$$N$, unitary matrices. For example, with
$9$ spatial dualizations, we have $9$ collective coordinates for
the $N$ coincident D0brane solitons: $A^{i}_{\mu}(x^0)$
$\leftrightarrow$ $X^{i}_{\mu} (x^0)$, $i$ $=$ $1$, $\cdots$, $N$,
and $\mu$ $=$ $1$, $\cdots$, $9$, where $x^0$ is time. Here, $i$
is the Chan-Paton index, and the gauge group realized on $N$
coincident D0branes is the nonabelian group $U(N)$, of rank $N$.
In the unoriented type I string theory we obtain, instead, the
orthogonal group $SO(2N)$ as worldvolume gauge group
\cite{polbook}.

\vskip 0.1in More generally, the $X^{\mu}$ coordinate location of
the $i$th D0brane is the $i$th eigenvalue of the $U(N)$ matrix
$X^i_{\mu}$, $i$$=$$1$, $\cdots$, $N$, described above.
Restricting to the Yang-Mills field theory on the one-dimensional
worldvolume of the D0branes, we have a worldvolume Lagrangian that
agrees precisely with the dimensional reduction of the 10d
nonabelian Yang-Mills Lagrangian. This gives the familiar quartic
interaction for one-dimensional $N$$\times$$N$ matrices
\cite{dlp,witnc}. Such matrix Hamiltonians describe the quantum
mechanics of, time-dependent, large $N$ unitary matrices, as in
the Banks-Fischler-Shenker-Susskind proposal for M(atrix) Theory
\cite{bfss}. Planar reduced matrix models, akin to the
Ishibashi-Kawai-Kitazawa-Tsuchiya IIB Matrix Model \cite{ikkt},
follow as the result of taking this logic one step further: we
must T-dualize all {\em ten} directions of spacetime. The
coordinates $X^i_{\mu}$, with $i$$=$$1$, $\cdots$, $N$, and $\mu$
$=$ $0$, $\cdots$, $9$, can now be interpreted as the locations of
$N$ {\em Dinstanton events} in a bulk ten-dimensional spacetime.
Recall that the tension of a Dinstanton has mass dimension zero.
Thus, such a matrix Lagrangian has no dimensionful couplings and
is reminiscent of a topological theory.

\vskip 0.1in It should be emphasized that the perspective we have
just described views the D(9-n)branes as fundamental degrees of
freedom in M theory, but also as semi-classical solitons that
exist in an $n$-dimensional classical flat spacetime geometry: an
underlying spacetime continuum has been assumed. Thus, the
M(atrix) theory conjecture of \cite{bfss} proposes an especially
simple Hamiltonian for an especially simple sector of M theory:
the remnant fundamental degrees of freedom that survive decoupling
in the infinite momentum frame. It should be noted that both
M(atrix) Theory \cite{bfss}, and the IIB Matrix Model \cite{ikkt},
are therefore conjectured theories of emergent, or induced,
linearized gravity: Newtonian gravity appears as an effective
long-distance interaction of fundamental, pointlike, degrees of
freedom, respectively, D0branes or Dinstantons, in an embedding
flat spacetime background. Reconstructing the full nonlinear
structure of the Einstein gravitational interaction from this
simplified starting point has proven prohibitively difficult
\cite{plefka}, as has the problem of extending the matrix model
formalism to backgrounds with curved spacetime geometries
\cite{taylor}. Although theories for an emergent gravitational
interaction, they are {\em not} fundamental theories of emergent
spacetime, and do not therefore capture the full spirit of
Einstein gravity. We have required more from our zero dimensional
matrix Lagrangian \cite{mat1}: we want both the emergence of a
continuum spacetime manifold, as well as the emergence of local
spacetime geometry. We wish to find this in a matrix Lagrangian
that can credibly encapsulate what we know about the Duality Web
of M theory, inclusive of all of its weakly coupled field and
string theory limits. Most importantly, the emergence of the
spacetime continuum is tied to
the large N limit. This spacetime geometry could genericallt receive
quantum corrections but, unlike the alternative proposals for
quantum gravity we comment upon below, quantum dynamics is {\em
not} an essential aspect of our proposal for an emergent continuum spacetime
geometry. The key point is that for renormalizable and anomaly-free
superstring theories with 16 supercharges, the flat spacetime background
is {\em exact}, without quantum back reaction.

\vskip 0.1in A brief clarification about alternative attempts, old
and new, to formulate quantum theories of gravity is in order.
Note that we are putting aside here the issue of whether such
formulations also extend to all of physics, namely, to the
Standard Model with its Yang-Mills gauge fields and chiral matter
multiplets, a unification which is one of perturbative superstring
theory's undeniable successes. Loop quantum gravity is perhaps the
best known attempt at writing a background independent theory of
quantum gravity in terms of alternate, loop variables: note that a
continuum spacetime manifold is assumed, but there is no spacetime
metric \cite{ashtekar}. Thus, this is {\em not} a theory of
emergent continuum spacetime, but neither does it contain a
satisfactory explanation for why one should trust a quantum field
theoretic formulation of 4d quantum general relativity given the
well-known non-renormalizability of perturbative gravity. The
general philosophy of identifying background independent variables
for a fundamental theory of emergent spacetime geometry is
nevertheless appealing, and the matrix formulation described in
this paper is a beautiful illustration of this phenomenon.

\vskip 0.1in Topological string theories, and topological M
theory, are theories of pform fields where a {\em topological}
gravity sector is conjectured to emerge from the dynamics of
p-form fields \cite{dijk,foam}. Notice that the loop
representation of 2+1 dimensional quantum gravity can incorporate
both the background spacetime manifold as well as its topological
degrees of freedom \cite{ashtekar}. But this property is special
to 2+1 dimensions. Spin network, and spin foam, models have also
been extensively explored as alternative formulations of quantum
general relativity \cite{baez}. A recent twist on these ideas are
the melting crystal configurations underlying the proposed notion
of {\em quantum foam} \cite{foam}, which also represents a
latticization of space. This is a rough analogue of the
discretized eigenvalue coordinates of the zero dimensional matrix
model. But the connections of the proposals in \cite{dijk,foam} to
genuine M theory, or even to genuine Einstein gravity, are at best
tenuous, and it is not clear to us how the observations in
\cite{foam,dijk} might extend to the full theory. They remain an
interesting exploration of certain special spacetime backgrounds
that can enter the quantum gravity path integral, representing
wild fluctuations in topology, and geometry, rather than smooth
spacetime metric geometries. Finally, neither can the AdS/CFT
conjecture \cite{malda} be considered the basis of a true theory
of emergent spacetime, except in the limited, holographic, sense
that the dynamics of a (D-1)-dimensional continuum quantum field
theory is being conjectured to describe all of physics in a
related D-dimensional continuum spacetime.

\vskip 0.1in Let us close with an outline of the paper. In section
2, we apply the simple procedure of planar reduction to a sample
supergravity Lagrangian with an auxiliary large $N$ flavor group,
and both with, and without, a Yang-Mills gauge sector. We present
our analysis using as prototype the manifestly supersymmetric 10d
Lagrangian density obtained in the low energy limit of the
heterotic string theory, computed up to quartic order in the
$\alpha^{\prime}$ expansion in \cite{br2,br}, and inclusive of
gauge-coupling dependent corrections required by closure of the
supersymmetry algebra. We compare the resulting planar reduced
matrix models with previously studied matrix models. Our
discussion includes an especially elegant and simple result for
the planar reduction of the 11d supergravity Lagrangian augmented
with the large N flavor symmetry. We explain why simple planar
reduction a la Eguchi-Kawai \cite{ek}, or the supersymmetric
extension in the IKKT matrix model \cite{ikkt}, always results in
the absence of any remnant of the spectrum of supergravity pform
potentials in the reduced matrix model, despite our introduction
of a large $N$ flavor symmetry in order to obtain a nontrivial
large N matrix model. We need a new idea.

\vskip 0.1in The introduction of a large N flavor index does not,
in itself, suffice as a device to obtain a matrix model Lagrangian
with remnant knowledge of the local spacetime symmetries of the
higher dimensional field theory. In section 3, we explain the
notion of reduction of spacetime fields to linearized forms on an
infinitesimal patch of local tangent space at a single point in
the spacetime manifold. The key new insight is to recognize that
the Lagrangian density in quantum field theory satisfies {\em
spacetime locality}. As a consequence, the reduction of all
spacetime fields to {\em linearized} forms on the infinitesimal
patch of local tangent space at a single point in the spacetime
manifold, suffices to preserve all of the local symmetries of a
covariant field theory Lagrangian in the corresponding reduced
matrix model. We also explain in section 3 the emergence of the
coordinates of a $D$-dimensional noncompact spacetime continuum in
the large N limit: a large N ground state of the matrix Lagrangian
which is characterized by $D$, simultaneously diagonalizable,
$N$$\times$$N$ matrices belonging to the zehn(elf)bein array,
$E_{\mu}^a d \xi^a$, and where $\xi^a$ parameterizes the
infinitesimal patch of local tangent space, will correspond to a
background of M theory with $D$ noncompact spacetime dimensions.
Such correspondences can be extended to generic curved spacetime
metrics. We also check in Section 3 the self-consistency of our
matrix Lagrangian with the basic relations of Riemannian geometry
in continuum spacetimes, with the form of infinitesimal Lorentz,
supersymmetry, and Yang-Mills, transformations, and with standard
spacetime field redefinitions, properties which are required to
emerge in the large N limit of the matrix model. The key aspect of
our prescription for spacetime reduction that enables such
self-consistency is the fact that the large $N$ flavor symmetry
group has been chosen to {\em commute} with the local symmetries
of the original field theory Lagrangian.

\vskip 0.1in In section 4, we describe our proposal for a
fundamental theory of emergent spacetime that also encapsulates
the insights of the string/M theory duality web, especially with regard 
to theories with sixteen supercharges \cite{chl,cp,mat1}.
Of special interest to us is the principle of generalized
electric-magnetic duality in the Dirichlet pbrane spectrum,
examined by us in the accompanying paper \cite{hodge}. We review
the intriguing fact that the full spectrum of supergravity pform
potentials, $-1$$\le$$p$$\le$$10$, is represented in the
nine-dimensional supergravity Lagrangian of \cite{berg}, up to
field redefinitions and dualities. Thus, the circle-compactified
type I-I$^{\prime}$-heterotic $E_8$$\times$$E_8$ and
Spin(32)/${\rm Z}_2$-IIA-IIB string theories appear in a single
spacetime Lagrangian, and on equal footing. An explicit analysis
of all of the two- and four-fermion terms in the nonmaximal 10D
N=1 supergravity-Yang-Mills Lagrangian required by closure of the
supersymmetry algebra, and up to quartic order in
$\alpha^{\prime}$ corrections, was performed by Bergshoeff and de
Roo in \cite{br2,br}, and is the case described in Section 3. The
analysis of the D=9 massive type IIA-IIB supergravity Lagrangian
described in \cite{berg} is far less comprehensive, restricted to
discussion of the bosonic sector alone, and with an absence of any
discussion of the $\alpha^{\prime}$ corrections from string
theory. In nine dimensions, there are no Poincare self-dual forms,
and it would be interesting to understand if a manifestly Poincare
dual form of the Lagrangian exists, exhibiting both electric and
magnetic dual p-form potentials \cite{cjlp2}. These issues are left to future
work. Instead, in sections 4 we address the complex nature of the
vacuum landscape of theories with sixteen supercharges: beyond the
standard component of the Moduli Space with rank 16 Yang-Mills
fields explored in the familiar \lq\lq Star" diagram, there are
disconnected components first discovered in the CHL analysis
\cite{chl,cp}. How are these theories accounted for in the matrix
model framework?

\vskip 0.1in How does our proposal relate to conjectures for the
hidden symmetry algebra of string/M theory? We address this
question in Section 5, emphasizing how the algebraic viewpoints
can lend significant insight into some key aspects of our proposal.  
In an accompanying paper \cite{matn6} we have given a
pedagogical review, presented from our own perspective, on recent
work on the hidden symmetry algebra of the ten and eleven
dimensional supergravities with 32 supercharges. Concrete evidence for generalized
electric-magnetic duality in the worldsheet formalism of
perturbative string theory has been given by us in
\cite{path,flux,hodge}, works done in partial collaboration with
Chen and Novak, and building on the earlier results in
\cite{cmnp}. The worldsheet computation of the tension of a
D(-2)brane coupling to a (-1)form supergravity potential, the
magnetic dual of the nine-form potential of massive IIA
supergravity, is described in the accompanying paper \cite{hodge},
along with a summary of the corroborating evidence for generalized
electric-magnetic duality presented by Schnakenburg and West in
their analysis of the global symmetry algebra of the massive IIA
supergravity \cite{w2}, and by the early brane spectrum analyses
by Lavrinenko, Lu, Pope, and Stelle \cite{llps}. West's arguments
in favor of the very-extended Lorentzian Kac-Moody algebra
$E_{11}$ as the hidden symmetry algebra of the ten and eleven
dimensional supergravities with 32 supercharges are summarized in
\cite{w4}; a review by us can be found in \cite{matn6}. The
relationship to our proposal for a fundamental theory of emergent
local spacetime geometry \cite{mat1}, and one that is also rooted
in the principle of generalized electric-magnetic duality as
presented in \cite{hodge}, is briefly described in sections 4 and
5. In particular, we conjecture a plausible extension of West's
$E_{11}$ framework which can incorporate theories with sixteen
supercharges, and a nontrivial Yang-Mills gauge sector, and which
is also compatible with the matrix model framework proposed by us
in \cite{mat1}. We conclude in Section 6 with a list of open
questions and some key directions for future work.

\section{Planar Reduction of Theories with Flavor $U(N)$}

\vskip 0.1in As explained above, there is no sign in either of the
previous matrix proposals for M theory \cite{bfss,ikkt} of the
global symmetries of the Einstein supergravity theories. It is
natural to suspect that the dimensional reductions of locally
supersymmetric Yang Mills theories would be a more relevant
direction to explore in this context and a concrete suggestion to
this effect was made by Nicolai in 1997-98 \cite{nicolai}. Notice,
however, that it is not immediately obvious why performing the
planar reduction of a {\em locally} supersymmetric large $N$
Yang-Mills theory is an interesting exercise: the standard
prescription for planar reduction \cite{ek} requires that we drop
all space and time derivatives, leaving a potential that is
quartic in the spin connection. But there are only a finite number
of degrees of freedom in the gravitational sector due to the
absence of large N symmetry: there is no room for interesting
quantum dynamics in the matrix model. Thus, it becomes necessary
to modify the Eguchi-Kawai prescription for planar reduction of
rigid Yang-Mills theories \cite{ek} when we study the
zero-dimensional reduction of a gravity theory.

\vskip 0.1in Let us begin with the ten-dimensional spacetime
Lagrangian of the anomaly-free heterotic supergravity coupled to
Yang-Mills fields \cite{fs,brwn,ch,br}. In the supergravity
literature, this is often referred to as the nonmaximal
ten-dimensional supergravity, in order to distinguish it from the
$N$$=$$2$ supergravities with 32 supercharges. The relevant
Lagrangian was constructed in several steps, starting with the
four-dimensional, $N$$=$$4$ supergravity multiplet covariantly
coupled to an abelian (Maxwell) multiplet \cite{fs}. This
Lagrangian can be straightforwardly interpreted as the dimensional
reduction of a ten-dimensional $N$$=$$1$ supergravity theory
coupled to an abelian gauge field \cite{brwn}. Extension to a
sector with nonabelian Yang-Mills fields requires inclusion of a
crucial Kalb-Ramond term in the Lagrangian, necessitated by the
mixed gauge-gravity anomaly cancellation conditions \cite{ch}.
Another feature of interest in this Lagrangian is the possibility
of a field-dualization: the Lagrangian can be expressed
interchangeably in terms of either a two-form, or a six-form,
supergravity potential \cite{ch,br}. The former choice is natural
from the perspective of the string supergravities: the low-energy
spacetime effective action of the heterotic string theory has a
Neveu-Schwarz sector two-form potential, $B_{ab}$, that can couple
to the fundamental closed string. In the low-energy spacetime
effective action of the type IB string theory, on the other hand,
the relevant two-form potential, $C_{ab}$, appears in the Ramond
sector, representing the possibility of coupling to a D1brane
\cite{polbook}.

\vskip 0.1in We have taken the liberty of simplifying the notation
and conventions given in the original references
\cite{fs,brwn,ch,br}. Our expressions correspond to the last of
these references.\footnote{Comparing with \cite{br2,br}, note the
following convenient replacements: $\phi$$\to$$e^{\Phi/3}$,
$\sqrtwo \lambda$$\to$$\lambda$, $B_{ab}$$\to$$\sqrtwo B_{ab}$,
$H_{abc}$$\to$$\sqrtwo H_{abc}$, $\beta$$\to$$g^2 \kappa^2$.
Finally, an overall factor of $-\half \kappa^2 E e^{-\Phi}$ has
been suppressed in writing Eq.\ (\ref{eq:hmat}).} Expressed in
terms of the Einstein frame metric, and to lowest order in the
inverse string tension, the $O(\alpha^{\prime 0})$ Lagrangian
takes the form \cite{br,mat1}:
\begin{eqnarray}
{\cal L} &&=~ \kap \left ( {\bar \psi}_{a} \Gamma^{abc} D_{b}
(\omega) \psi_{c} ~-~ 4 {\bar{\lambda}} \Gamma^{ab} D_{a} (\omega)
\psi_{b} ~-~ 4 {\bar{\lambda}} \Gamma^{a} D_{a} (\omega) \lambda
\right ) ~+~ \betag ~ {\bar \chi}^i \Gamma^{a} D_{a} (\omega,A)
\chi^i \cr \quad\quad &&\quad\quad  ~+~ \kap \left ( {\cal
R}(\omega,E) ~-~  \partial^{a} \Phi ~\partial_{a} \Phi ~+~ 3
{H}^{abc} {H}_{abc} \right )~+~ \half \betag ~ {F}^{ab}(A) F_{ab}
(A) \cr \quad && \quad\quad\quad\quad ~+~ {\cal L}_{\rm 2-fermi}
~+~ {\cal L}_{\rm 4-fermi}  \quad . \label{eq:hmat}
\end{eqnarray}
Here, $\psi_a$, $\lambda$, and $\chi$, denote, respectively, the
gravitino, dilatino, and gaugino, of the $N$$=$$1$, $D$$=$$10$,
supergravity-Yang-Mills theory. The covariant derivative, $D_a$,
is covariantized with respect to both Lorentz and Yang-Mills gauge
transformations. ${\cal R}$ is the ten-dimensional Einstein
curvature scalar, and $\Phi$ is the dilaton. We work in Palatini's
first-order formulation for Einstein gravity with a zehnbein,
$E^{\mu}_a$, and spin-connection, $\omega_{\mu}^{ab}$. $F_{ab}$
and $H_{abc}$ are, respectively, the two-form Yang-Mills and
three-form supergravity field strength. The $\{ \Gamma^a \}$ are
the rank sixteen Dirac matrices of the ten-dimensional Clifford
algebra; we have suppressed all spinor indices. Finally, $g$ and
$\kappa$ denote, respectively, the dimensionless part of the
physical, ten-dimensional Yang-Mills and gravitational couplings.
In addition, we must include the crucial two-fermion and
four-fermion terms required by supersymmetry \cite{brwn,br}. The
two-fermi terms take the form:
\begin{eqnarray}
 {\cal L}_{\rm 2-fermi} &&=
 {\bar{\psi}}_{a} \Gamma^{a} \psi_{b} \partial^{b} \Phi ~-~ 2 ~{\bar
{\psi}}_{a} \Gamma^{b} \Gamma^{a} \lambda \partial_{b} \Phi \cr
&&\quad\quad - \quart H^{def} \left [ {\bar \psi}_{a} \Gamma^{[a}
\Gamma_{def }\Gamma^{b]} \psi_{b} ~+~ 4 ~{\bar \psi}_{a}
\Gamma^{a}_{def } \lambda ~-~ 4 ~ {\bar\lambda} \Gamma_{def }
\lambda ~+~ \betag {\bar \chi} \Gamma_{def} \chi \right ] \cr
&&\quad\quad\quad ~+~ \quart \betag ~ {\bar \chi}^i \Gamma^{d}
\Gamma^{ab} (\psi_{d} +  \thre \Gamma_{d} \lambda ) (F_{ab}^i +
{\hat{F}}_{ab}^i ) \quad . \label{eq:2fermat}
\end{eqnarray}
Likewise, the 4-fermi terms in the ten-dimensional covariant
supergravity-Yang-Mills Lagrangian take the form:
\begin{eqnarray}
{\cal L}_{\rm 4-fermi} &&=  \ninsix {\bar\psi}^{f} \Gamma^{abc}
\psi_{f}
 \left ( \half {\bar\psi}_{d} \Gamma^{d} \Gamma_{abc} \Gamma^{e} \psi_{e}
      + ~ {\bar{\psi}}^{d} \Gamma_{abc} \psi_{d}
        - 2 ~ {\bar{\lambda}} \Gamma_{abc} \lambda
          - 2~ {\bar{\lambda}} \Gamma_{abc} \Gamma^{d}\psi_{d}
                   \right )
\cr && \quad - \ninsix \betag {\bar{\chi}}^i \Gamma^{abc} \chi^i
\left ( {\bar{\psi}}_{d} ( 4 \Gamma_{abc} \Gamma^{d} + 3
\Gamma^{d} \Gamma_{abc} ) \lambda
   +2 {\bar{\lambda}}\Gamma_{abc}\lambda + 3\cdot 2^3 H_{abc} + {\bar{\chi}}^j \Gamma^{abc}
\chi^j \right ) . \label{eq:4fermimat}
\end{eqnarray}
The Einstein curvature, Yang-Mills, and three-form field strengths
take the familiar form:
\begin{eqnarray}
{\cal R} =
\partial_{a} \omega_{b}^{ab} -
\partial_{b} \omega_{a}^{ab} +
\omega_{b}^{ac} \omega_{a c}^b  - \omega_{a}^{ac} \omega_{b c}^b
 \cr F^i_{ab}  =
\partial_{a} A^i_{b} - \partial_{b} A^i_{a} +
f^{ijk} A^j_{a} A^k_{b}  \cr H_{abc} =
\partial_{[a} B_{bc]} - \betag
 (A^i_{[a}\partial_{b} A^i_{c]}
  - {{2}\over{3}} f^{ijk} A^i_{[a} A^j_{b} A^k_{c ]} )
\quad . \label{eq:fields}
\end{eqnarray}
Here, $[ \tau^i , \tau^j ] $$=$$i f^{ijk} \tau^k$, defines the
structure constants, $f^{ijk}$, of the Yang-Mills gauge group,
$G$, with $i,j,k$ $=$ $1$, $\cdots$, ${\rm dim}~G$. Notice the
Chern-Simons contribution to the three-form field strength. Hats
on the curvatures denote the supercovariant derivatives
\cite{brwn,br}. Explicitly, we have \cite{br2,br}:
\begin{eqnarray}
{\hat{D}}_a \Phi =&& \partial_a \Phi + \sqrtwo {\bar{\psi}}_a
\lambda \cr
 {\hat{D}}_a (\omega) \lambda =&& D_a (\omega) \lambda
+ \eigth \sqrtwo \Gamma^b \psi_a D_b \Phi - \eigth \Gamma^{abc}
\psi_{a} \left ( {\hat{H}}_{abc} - \to {\bar{\lambda}}
\Gamma_{abc}
 \lambda \right ) - {{1}\over{3 \cdot 2^7}} \betag \sqrtwo \Gamma^{bcd}
 \psi_{a} {\bar{\chi}}^i \Gamma_{bcd} \chi^i \cr
 {\hat{D}}_a (\omega, A) \chi^i =&& D_a (\omega , A) \chi^i +
 \quart \Gamma^{bc} \psi_a {\hat{F}}_{bc} - \half \left ( \psi_a
 {\bar{\chi}}^i \lambda - \chi^i {\bar{\psi}}_a \lambda + \Gamma^b
 \lambda {\bar{\chi}}^i \Gamma_b \psi_a \right )
\quad , \label{eq:covd}
\end{eqnarray}
and the supercovariantized field strengths take the form:
\begin{eqnarray}
{\hat{F}}^i_{ab} =&& F^i_{ab} - {\bar{\psi}}_{[a} \Gamma_{b]}
\chi^i \cr {\hat{H}}_{abc} =&& \partial_{[a} B_{bc]} - \quart
{\bar{\psi}}_{[a} \Gamma_b \psi_{c]} - \betag \sqrtwo \left \{
A^i_{[a}
\partial_b A^i_{c]} - \third f_{ijk} A^i_{[a} A^j_{b} A^k_{c]}
\right \} \quad . \label{eq:fieldss}
\end{eqnarray}
We should reiterate that the Lagrangian described above is,
strictly speaking, that of the heterotic 10d
supergravity-Yang-Mills string supergravity \cite{br}. The
two-form potential $B_{ab}$ couples to fundamental heterotic
closed strings. Now consider introducing a {\em flavor} quantum
number in the Einstein-Yang-Mills Lagrangian given in Eq.\
(\ref{eq:hmat}), replacing the gravitational spin-connection and
Yang-Mills vector potential with $N$$\times$$N$ arrays as follows:
\begin{eqnarray}
{\cal R} \to&&  \partial_{a} (\omega_{b}^{ab})_{AB} -
\partial_{b} (\omega_{a}^{ab})_{AB} + (\omega_{b}^{ac})_{AC}
(\omega_{a c}^b)_{CB}  - (\omega_{a}^{ac})_{AC} (\omega_{b
c}^b)_{CB}  \cr F^i_{ab} \to&&
\partial_{a} (A^i_{b})_{AB} - \partial_{b} (A^i_{a})_{ AB} +
f^{ijk} (A^j_{a})_{ AC } (A^k_{b})_{ CB}  \quad ,
\label{eq:ymlarge}
\end{eqnarray}
where the indices run from $ i$$=$$1$, $\cdots$, ${\rm dim}~ G$,
and $A,B$$=$$1$, $\cdots$, $ N$. We will need to include a trace
over the $U(N)$ flavor group in order that the new Lagrangian
density transform as a $U(N)$ singlet. This Lagrangian will have
the symmetry group $U(N)$$\times$$G$, except that the $U(N)$ is
{\em not} a gauge symmetry. Rather, it plays the role of a flavor
group. How does one give meaning to the planar reduction of a
locally supersymmetric Lagrangian with a {\em huge} flavor
symmetry group to a single spacetime point? And why have we
introduced a large $N$ flavor symmetry, as opposed to the usual
large $N$ gauge symmetries invoked in \cite{ek,bfss,ikkt}?

\vskip 0.1in Notice that if we were to carry out the large $N$
extension in analogy with the planar reductions of rigid
Yang-Mills Lagrangians \cite{ek,ikkt}, namely, replace the
anomaly-free Yang-Mills group with the unitary large $N$ group:
$SO(32)$ $\to$ $U(N)$, where $SO(32)$ $\subset$ $U(32)$ $\subset$
$U(N)$, and where $U(N)$ is a fully {\em gauged} symmetry, we
would find nothing of interest in the gravity sector of the
Lagrangian. Suppressing all spacetime derivatives in the
supergravity-Yang-Mills Lagrangian as usual, we obtain the
standard quartic unitary matrix potential in the Yang-Mills
sector: $[A_{\mu},A_{\nu}]^2$, where $A$ is now a matrix of rank
$N$, but the Einstein sector yields an uninteresting finite rank
correction to these terms. Thus, since the zehnbein and spin
connection are {\em finite}-dimensional matrix arrays when reduced
to a single spacetime point, dimensional reduction of the Einstein
action gives a {\em finite} number of terms of the general form,
$E^{\mu}_a \omega^{ac}_{\mu} \omega^{b}_{c\lambda} E^{\lambda}_b
$. It is clear that if one desires a nontrivial modification of
large $N$ dynamics of the Yang-Mills sector {\em the gravity
variables must also scale with $N$}. This implies that the role of
$U(N)$ in the continuum field theory Lagrangian must be that of a
{\em flavor} symmetry group, rather than of a gauge
group.\footnote{I would like to thank Bernard de Witt for seeking
this clarification.} We will find that the introduction of $U(N)$
as a flavor symmetry in the continuum Lagrangian enables the
democratic appearance of large $N$ scaling behavior in both the
gauge and gravity sectors of the matrix model obtained upon
spacetime reduction to a single point.

\vskip 0.1in This motivates the first innovation introduced by us
in \cite{mat1}: both the vector potential, zehnbein, and spin
connection, were required to transform in the adjoint
representation of a large $N$ flavor group. The same requirement
was made of the other bosonic fields in the Lagrangian, namely,
the dilaton and two-form potential. What about the spinors in the
supergravity Lagrangian? Since we have required that the large $N$
flavor group commute with the group of supersymmetry
transformations, it is important that fields which are partners
under supersymmetry belong to the same $U(N)$ representation.
Thus, we will require that all spinor fields gravitino, dilatino,
and gaugino, also transform in the adjoint representation of the
large $N$ flavor group. Keeping only the terms in Eq.\
({\ref{eq:hmat}) that remain after planar reduction, gives the
following supermatrix Lagrangian:\footnote{We denote the spacetime
field, $f(x)$, and its planar-reduced representative which lives
at the origin of spacetime, $f(0)$, by the same symbol $f$.}
\begin{eqnarray}
{\cal L}^{(10d)}_{\rm planar} &&=~ \half \gap {\rm tr} \left (
f^{ijk} f^{ilm} A^{l a} A^{m b} A^j_{a} A^k_{b} ~+~ {\bar \chi}^i
\Gamma^a A_{a}^{j} \tau^{j} \chi^i \right ) \cr \quad\quad &&\quad
~+ \kap {\rm tr} \left ( E_a^{\mu} \left [ \omega_{\nu}^{ac}
\omega_{\mu c}^b  - \omega_{\mu}^{ac} \omega_{\nu c}^b \right ]
E_{b}^{\nu} - A^i_{a} \tau^i \Phi A_{a}^j \tau^j \Phi  + {\bar
\chi}^i \Gamma^a \omega_{abc}\Gamma^{bc} \chi^i \right ) \cr \quad
&&\quad\quad + \kap {\rm tr} \left ( {\bar \psi}_{a} \Gamma^{abc}
\omega_{bde}\Gamma^{de} \psi_{c} - 4 {\bar{\lambda}} \Gamma^{ab}
\omega_{ade}\Gamma^{de} \psi_{b} - 4 {\bar{\lambda}} \Gamma^{a}
\omega_{ade}\Gamma^{de} \lambda \right ) \cr \quad &&
\quad\quad\quad + {{4}\over{3}} {{1}\over{g^4}} {\rm tr} \left (
f^{nlm} f^{ijk} A^{n[a} A^{l b} A^{m c ]} A^i_{[a} A^j_{b} A^k_{c
]} \right )
   + {\rm tr} ~{\cal L}_{\rm 2-fermi} ~+~ {\rm tr} ~{\cal L}_{\rm
4-fermi}  \quad . \label{eq:lmat}
\end{eqnarray}
where $i,j,k, \cdots $ are group indices for the
finite-dimensional Yang-Mills gauge group, and repeated indices
are to be summed. The notation \lq\lq tr" denotes, instead, the
trace over the large $N$ flavor group, whose indices have been
suppressed. The first line of this expression is familiar:
analogous terms appear in both the Banks-Fischler-Shenker-Susskind
\cite{bfss} and Ishibashi-Kawai-Kitazawa-Tsuchiya \cite{ikkt}
rigid matrix models. The index structure of the terms in the first
line make the $U(N)$$\times$$G$ symmetry of the supermatrix model
manifest: the model obtained by restricting to only the terms in
the first line of this expression defines the {\em simplest}
possible supermatrix model consistent with this symmetry group.

\vskip 0.1in Thus, the distinction between flavor, and gauged,
large $N$ symmetry becomes significant only when we take into
account the remaining terms in the matrix Lagrangian: we find new
large $N$ matrix variables originating in the supergravity sector
of the continuum Lagrangian, as well as new multi-matrix
interaction terms. These include a sixth-order self-interaction
for the Yang-Mills potential, a term which was absent in both the
BFSS and IKKT matrix models \cite{bfss,ikkt}, and which arises
from the Chern-Simons contribution to the supergravity three-form
field strength. It is evident that the symmetry structure of the
full supermatrix Lagrangian given in Eq.\ (\ref{eq:lmat}) is much
more subtle than simply $U(N)$$\times$$G$. Knowledge of the
Cremmer-Julia hidden symmetries of the continuum field theory
Lagrangian becomes a useful tool for its analysis.

\vskip 0.1in It is illuminating to examine the form of matrix
Lagrangian obtained by the planar reduction of the 11d
supergravity theory. Recall the absence of Yang-Mills gauge
fields, as well as the absence of a dilaton supermultiplet, in 11d
supergravity. The 11d supergravity theory does, however, include a
four-form field strength. The associated three-form potential
couples to the supermembrane. Introduction of a large $N$ flavor
quantum number in the continuum Lagrangian, followed by spacetime
reduction of the field theory to a single spacetime point, gives
an elegant and especially simple matrix model:
\begin{eqnarray}
{\cal L}^{(11d)}_{\rm planar} =&& \kap {\rm tr} \left \{ {\bar
\psi}_{a} \Gamma^{abc} \omega_{bde} \Gamma^{de} \psi_{c} +
\omega_{b}^{ac} \omega_{a c}^b - \omega_{a}^{ac} \omega_{b c}^b
\right \} \quad , \label{eq:11dmat}
\end{eqnarray}
where the gravitino, $\psi_a$, is a 32-component Grassmann-valued
array, and $\omega_{abc}$ is the spin-connection. Notice that the
the presence of a higher p-form supergravity potential in the
continuum field theory Lagrangian is, unfortunately, erased from
the planar reduced matrix model: there is no analogous
Chern-Simons coupling to a Yang-Mills field, as was present in
Eq.\ (\ref{eq:lmat}). It may well be true that this particular
supermatrix Lagrangian falls within the class of solvable
zero-dimensional multi-matrix models, enabling a detailed analysis
by well-established matrix model techniques. We should emphasize
that this model is the precise, pure gravitational, supermatrix
model analog of the planar reductions of rigid supersymmetric
Yang-Mills theory considered in \cite{ek,bfss,ikkt}. However, the
model appears not to capture the full content of M theory because
it lacks any knowledge of the crucial supermembrane sector of the
theory.

\vskip 0.1in Thus, while the planar reduction of gravity theories
with large $N$ flavor group has led to an interesting new class of
zero-dimensional matrix models, these models appear not to capture
the full content of M theory, inclusive of the crucial
brane-spectrum required by duality. This brings us to a second
innovation introduced in \cite{mat1}. Notice that, strictly
speaking, the Lagrangian in Eq.\ (\ref{eq:hmat}) describes a
ten-dimensional supergravity theory in generic curved spacetime
background. Inherent in this expression is the notion of a local
ten-dimensional flat tangent space attached to every point in
spacetime. The naive procedure of planar reduction we have
borrowed from rigid super-Yang-Mills theories \cite{ek} has
ignored this aspect of the supergravity Lagrangian. We will now
show that the spacetime reduction of all spacetime fields to {\em
linearized} forms defined on the infinitesimal patch of local
tangent space at a single spacetime point, suffices to ensure that
all of the local symmetries of the continuum Lagrangian are
preserved in a corresponding reduced matrix model.

\section{Emergence of the Spacetime Continuum}

\vskip 0.1in We proceed with formulating a prescription for the
spacetime reduction of a $d$-dimensional supergravity theory
coupled to Yang-Mills fields to a single spacetime point together
with an infinitesimal patch of a local flat tangent space. As
emphasized above, our starting point is an unusual field theory
Lagrangian with a huge flavor symmetry group: all fields, bosonic
and fermionic, are required, in addition, to live in the adjoint
representation of the large $N$ unitary group, $U(N)$. We
emphasize that $U(N)$ is a flavor symmetry; only the finite rank
anomaly-free Yang-Mills group $G$ has been gauged. Starting with
the nonmaximal d=10 supergravity theory coupled to $O(32)$
Yang-Mills fields, we have the corresponding $U(N)$ invariant
Lagrangian:
\begin{eqnarray}
{\cal L} &&=~ \kap {\rm tr} \left \{ {\bar \psi}_{a} \Gamma^{abc}
D_{b} (\omega) \psi_{c} ~-~ 4 {\bar{\lambda}} \Gamma^{ab} D_{a}
(\omega) \psi_{b} ~-~ 4 {\bar{\lambda}} \Gamma^{a} D_{a} (\omega)
\lambda \right \} \cr \quad &&\quad +~ \gap ~ {\rm tr} \left \{
{\bar \chi}^i \Gamma^{a} D_{a} (\omega,A) \chi^i~+~ \half ~
{F}^{ab}(A) F_{ab} (A) \right \} \cr \quad\quad &&\quad\quad ~+~
\kap {\rm tr} \left \{ {\cal R}(\omega,E) ~-~
\partial^{a} \Phi ~\partial_{a} \Phi ~+~ 3 {H}^{abc} {H}_{abc}
\right \} \cr && \quad\quad\quad\quad~+~ {\rm tr} \left \{ {\cal
L}_{\rm 2-fermi} ~+~ {\cal L}_{\rm 4-fermi}  \right \} \quad .
\label{eq:hmatflav}
\end{eqnarray}
where the notation \lq\lq tr" denotes taking the trace over the
large $N$ flavor group, and the two- and four-fermi terms are as
given by Eqs.\ (\ref{eq:2fermat}) and (\ref{eq:4fermimat}). Notice
that each term in the Lagrangian is a flavor singlet, and the
$U(N)$ flavor group commutes with {\em all} of the spacetime
symmetries of the Lagrangian: namely, local Lorentz and local
supersymmetry transformations, in addition to Yang-Mills gauge
transformations.\footnote{In our earlier papers \cite{mat1}, we
have pointed out a more general possibility for the matrix
superalgebra. Namely, the parameters for infinitesimal
supersymmetry and $SL(n,{\bf R})$ transformations could themselves
be non-singlet under the flavor $U(N)$. While we know of no reason
to rule out such an extension, it is not necessary for the problem
at hand. Notice that, for such matrix algebras, the large $N$
limit would have to correspond to an exotic (nonlinear) extension
of the Nahm classification of spacetime linear superalgebras. We
thank Bernard de Witt for pointing this out.}

\vskip 0.1in The $E_a^{\mu}$ are the fundamental variables
appearing in the matrix Lagrangian, but they are not all
independent. Assuming a flat tangent space of Minkowskian
signature, $\eta_{ab}$, the usual relation for the spacetime
metric tensor takes the form of a $U(N)$ identity:
\begin{equation}
G^{\mu\nu} = {\rm tr} \left ( E^{\mu}_a E^{\nu}_b \right ) , \quad
\mu,\nu, ~ {\rm and} ~ a,b=0, \cdots 9 , \quad\quad E^{\mu}_a =
G^{\mu\nu} E_{\nu a} \quad . \label{eq:basicsc}
\end{equation}
As is familiar from differential geometry, $G^{\mu\nu}$ is the
object that raises spacetime indices, while $\eta^{ab}$ is the
object that raises indices in tangent space. The spacetime metric
transforms as a $U(N)$ singlet, as does $\eta^{ab}$. The usual
constraint equation relating them is automatically satisfied:
\begin{equation}
\eta_{ab} = {\rm tr} \left ( E^{\mu}_{ a} E_{\mu b} \right ) =
G^{\mu\nu} {\rm tr} \left ( E_{\nu a } E_{\mu b} \right ) =
G^{\mu\nu} {\rm tr} \left ( G_{\nu \lambda} \eta_{ac} E^{\lambda
c} E_{\mu b} \right ) = \delta^{\mu}_{\lambda} {\rm tr} \left (
\eta_{ac} E^{\lambda c} E_{\mu b} \right ) = \eta_{ab} \quad .
\label{eq:constrmet}
\end{equation}

\vskip 0.1in Consider the dimensional reduction of the large $N$
Lagrangian of Eq.\ (\ref{eq:hmatflav}) to a single spacetime
point, call this the origin of 10d spacetime, {\em together with
an infinitesimal patch of the local flat tangent space}. In other
words, instead of simply setting all spacetime derivatives to zero
as in the previous section, we retain the $O(\delta \xi^a)$ terms
of the continuum Lagrangian, truncating at $O((\delta \xi^a)^2)$
in the Taylor expansion on tangent space, in the infinitesimal
vicinity of the spacetime origin. We have parameterized the
infinitesimal patch of local tangent space at the origin by the
variables $\xi^a $, $a$$=$$0$, $\cdots$, $9$. Recall the usual
relation in Riemannian differential geometry linking the partial
derivative operators acting in spacetime, and in the local tangent
space:
\begin{equation}
\partial_{\mu} = E_{\mu}^a \partial_a , \quad \quad \mu , \nu = 0, \cdots , 9,
\quad a,b = 0, \cdots , 9 \quad , \label{eq:basics}
\end{equation}
Since the zehnbein is a flavor adjoint, an $N$$\times$$N$
dimensional array, whereas $\partial/\partial \xi^a$ is the
ordinary partial derivative operator acting on a continuous and
differentiable space with the local geometry of $R^{10}$, {\em it
follows that the partial derivative operator in spacetime,
$\partial_{\mu}$, is also $U(N)$ valued}. In particular,
consistency with the obvious identity $\partial_{\mu}
X^{\mu}$$=$$1$, implies that:
\begin{equation}
( X^{\mu})_{AB} \equiv (E^{\mu}_{ a})_{AB} \delta \xi^a , \quad
\quad (\partial_{\mu})_{AB} (X^{\mu})_{BC} = ({\bf 1})_{AC} ,
\quad \quad A,B =1, \cdots N \quad . \label{eq:vielb}
\end{equation}
In other words, the coordinate vector, $X^{\mu}$, is itself $U(N)$
valued! Notice that $X^{\mu}$ is a dependent variable in our
framework: it is {\em derived} from the zehnbein, $E_{\mu}^a$,
which is the fundamental variable appearing in the matrix
Lagrangian. Unlike tangent space, which is smooth and
differentiable, at least infinitesimally, spacetime contains a
single element, a single spacetime \lq\lq point". All variables
defined at a single point of the spacetime manifold are
$N$$\times$$N$ dimensional matrices, the fundamental degrees of
freedom in the matrix model Lagrangian. Further, we will require
of all pforms on tangent space that they satisfy the {\em
linearized} property: while the partial derivatives at the origin,
$\partial_n f(0)$ $=$ $0$, $n\ge2$, exist up to arbitrarily order,
the higher-order terms in the Taylor expansion are absent because
quadratic and higher powers of the infinitesimal, $\delta \xi^a$,
in tangent space has been dropped, reflecting the fact that we
have a set of pforms on a base manifold of {\em infinitesimal}
extent.

\vskip 0.1in Why is there a need to retain an infinitesimal patch
of tangent space while performing the dimensional reduction of the
gravity theory to a single spacetime point? To develop some
intuition into the $U(N)$ valued relations given above, notice
that no restrictions have been placed upon the eigenvalue spectrum
of the various zehnbein. In principle, one can solve for the
eigenvalue spectrum of each $E^{\mu}_a$, given the equation of
motion that follows from the classical matrix Lagrangian. One of
the solutions to the equation of motion corresponds to choosing
the 10d Minkowskian flat space time metric as classical
background:
\begin{equation}
<G^{\mu\nu}> ~=~ <{\rm tr} \left ( E^{\mu}_a E^{\nu}_b \right )>
~=~ \eta^{\mu\nu} , \quad \mu,\nu, ~ {\rm and} ~ a,b=0, \cdots 9
\quad . \label{eq:basicflat}
\end{equation}
We solve for the corresponding $<E_{\mu}^a>$, expressing them in
diagonal form, and ordering the eigenvalues along the diagonal to
reflect a {\em monotonic increase}. It is evident that in the
large $N$ limit, the eigenvalues will crowd together forming a
continuum. Of course, as a consequence of the identity in Eq.\
(\ref{eq:vielb}), the coordinate matrices, $<X^{\mu}>$, also take
diagonal form, their entries reflecting the monotonic increase
along the diagonals of individual zehnbein. It is natural to
interpret the ordered continuum of eigenvalues of the coordinate
matrix as coordinate-locations for the continuum of spacetime
points along the coordinate axis $x^{\mu}$ of 10d Minkowskian
spacetime. Thus, we have recovered the coordinates of the
spacetime continuum by taking the large $N$ limit of the matrix
model!

\vskip 0.1in  We are now ready to carry out the spacetime
reduction of the Lagrangian given in Eq.\ (\ref{eq:hmatflav}) in
accordance with our new prescription. Note that all fields,
bosonic or fermionic, transform as adjoints under the flavor
$U(N)$, and every term in the Lagrangian is a $U(N)$ singlet. The
Lagrangian is manifestly invariant under local supersymmetry and
local Lorentz transformations, and these symmetries commute with
the flavor $U(N)$. Under spacetime reduction, every field in the
Lagrangian is reduced to a {\em linearized} pform on the local
tangent space, reflecting the fact that tangent space is an {\em
infinitesimal} manifold. Most importantly, this also has the
natural consequence that the local symmetries of the continuum
Lagrangian can be made manifest in the matrix Lagrangian.

\vskip 0.1in The remaining independent dynamical fields in the
Lagrangian reduce to {\em linear} forms on tangent space. Our
notation for a generic linearized form $f(\xi)$ is as follows:
$f(\xi)$ $=$ $f(0)$ $+$ $\partial_a f(0)$ $ \delta \xi^{a}$, where
$\partial_a f(0)$ denotes, more precisely, the partial derivative
of $f$ with respect to $\xi^a$, evaluated at $\xi^a$ $=$ $0$.
Since every field in the continuum Lagrangian is an $N$$\times$$N$
array under flavor $U(N)$, and global symmetries are preserved
under spacetime reduction, $f(0)$ and $\partial_a f(0)$ are two
{\em independent} unitary matrices appearing in the matrix
Lagrangian. Of course, one or other matrix array will be found to
drop out of any given term in the Lagrangian. Thus, the matrix
Lagrangian will turn out to have {\em exactly} the same symmetry
group as the original continuum field theory Lagrangian. Listing
each of the independent dynamical fields appearing in Eq.\
(\ref{eq:hmatflav}), we have the following result upon spacetime
reduction to corresponding $N$$\times$$N$ matrix arrays defined at
the origin $x$ $=$ $0$, which we choose coincident with the origin
of tangent space, $\xi$ $=$ $0$:
\begin{eqnarray}
E^{\mu}_a (x) &&\to E^{\mu}_a (0)  \cr A_a^i (x) &&\to A_a^i(0) +
\partial_b A_{a}^i(0) \delta \xi^b  \cr
 \partial_{c} A_{d}^i (x)  &&\to \partial_c (A_d^i(0)) + \partial_b A_{d}^i (0) \partial_c ( \delta \xi^b )
  = \partial_c A_{d}^i (0) \cr
D_c A_d^i (x) &&\to \partial_c A_{d}^i(0)  - \partial_d A_{c}^i
(0) + f^{ijk} A_c^j(0) A_d^k(0) \cr
 H_{abc} (x) &&\to
\partial_{[c} B_{ab]} (0) - g^2\left ( A^i_{[a} (0)
\partial_{b} A^i_{c]} (0)
  - \twothird f^{ijk} A^i_{[a}(0) A^j_{b}(0) A^k_{c ]}(0) \right ) \cr
  D_a \Phi (x) && \to \partial_a \Phi (0) - A_a^i \tau^i \Phi(0)
\quad , \label{eq:formsf}
\end{eqnarray}
where the indices have range as follows: $\mu $ $=$ $0$, $\cdots$,
$9$, $a,b,c$ $=$ $0$, $\cdots$, $9$, and $ i,j,k$ $=$ $1$,
$\cdots$, ${\rm dim} ~ G$. We remind the reader that each of the
objects on the left-hand-side of this list is also an
$N$$\times$$N$ unitary matrix, the flavor indices have simply been
suppressed. Suppressing the \lq\lq$(0)$" dependence, we obtain the
matrix Lagrangian:
\begin{eqnarray}
{\cal L}^{\rm (mat)} &&=~ \kap {\rm tr} \left \{ {\bar \psi}_{a}
\Gamma^{abc} D_{b} (\omega) \psi_{c} ~-~ 4 {\bar{\lambda}}
\Gamma^{ab} D_{a} (\omega) \psi_{b} ~-~ 4 {\bar{\lambda}}
\Gamma^{a} D_{a} (\omega) \lambda \right \} \cr && \quad ~+~ \kap
{\rm tr} \left \{ {\cal R}(\omega,E) ~-~ \partial^{a} \Phi
~\partial_{a} \Phi ~+~ 3 {H}^{abc} {H}_{abc} \right \}
\cr&&\quad\quad ~+~ \gap {\rm tr} \left \{ \half ~ {F}^{ab}(A)
F_{ab} (A) + ~ {\bar \chi}^i \Gamma^{a} D_{a} (\omega,A) \chi^i
\right \} \cr \quad && \quad\quad\quad\quad ~+~ {\cal L}^{\rm
(mat)}_{\rm 2-fermi} ~+~ {\cal L}^{\rm (mat)}_{\rm 4-fermi}  \quad
. \label{eq:hmatfm}
\end{eqnarray}
In other words, the matrix Lagrangian takes precisely the {\em
same} form as the original continuum Lagrangian with large $N$
flavor group, except that all spacetime fields are restricted to
their value at the origin: the infinite number of degrees of
freedom in the original continuum field theory have indeed been
drastically thinned to those of a zero-dimensional matrix model
with $U(N)$ flavor symmetry. But, remarkably, by the introduction
of linearized forms on the local flat tangent space, this matrix
Lagrangian also preserves a remnant of the {\em local} symmetries
of the continuum Lagrangian. The underlying reason why there
exists a matrix Lagrangian that can make manifest the local
symmetries of a given continuum field theory, is the spacetime
{\em locality} property of the Lagrangian density in quantum field
theory.

\vskip 0.1in For completeness, let us reproduce the crucial
two-fermion, and four-fermion, terms required by supersymmetry
\cite{brwn,br} for the matrix Lagrangian. The two-fermi terms take
the form:
\begin{eqnarray}
 {\cal L}^{\rm (mat)}_{\rm 2-fermi} &&=
 {\rm tr} \left \{ {\bar{\psi}}_{a} \Gamma^{a} \psi_{b} \partial^b \Phi ~-~ 2 ~{\bar
{\psi}}_{a} \Gamma^{b} \Gamma^{a} \lambda \partial_{b} \Phi \right
\} \cr &&\quad\quad - \quart {\rm tr} \left \{  H^{def} \left [
{\bar \psi}_{a} \Gamma^{[a} \Gamma_{def }\Gamma^{b]} \psi_{b} ~+~
4 ~{\bar \psi}_{a} \Gamma^{a}_{def } \lambda ~-~ 4 ~ {\bar\lambda}
\Gamma_{def } \lambda ~+~ \gap  {\bar \chi} \Gamma_{def} \chi
\right ] \right \} \cr &&\quad\quad\quad ~+~ \quart g^2 ~{\rm tr}
\left \{ {\bar \chi}^i \Gamma^{d} \Gamma^{ab} (\psi_{d} +  \thre
\Gamma_{d} \lambda ) (F_{ab}^i + {\hat{F}}_{ab}^i ) \right \}
\quad . \label{eq:2fermatm}
\end{eqnarray}
Likewise, the 4-fermi terms take the form:
\begin{eqnarray}
{\cal L}^{\rm (mat)}_{\rm 4-fermi} &&=  \ninsix {\rm tr} \left \{
{\bar\psi}^{f} \Gamma^{abc} \psi_{f}
 \left ( \half {\bar\psi}_{d} \Gamma^{d} \Gamma_{abc} \Gamma^{e} \psi_{e}
      + ~ {\bar{\psi}}^{d} \Gamma_{abc} \psi_{d}
        - 2 ~ {\bar{\lambda}} \Gamma_{abc} \lambda
          - 2~ {\bar{\lambda}} \Gamma_{abc} \Gamma^{d}\psi_{d}
                   \right ) \right \}
\cr &&  - \ninsix \gap {\rm tr} \left \{ {\bar{\chi}}^i
\Gamma^{abc} \chi^i \left ( {\bar{\psi}}_{d} ( 4 \Gamma_{abc}
\Gamma^{d} + 3 \Gamma^{d} \Gamma_{abc} ) \lambda
   +2 {\bar{\lambda}}\Gamma_{abc}\lambda + 3\cdot 2^3 H_{abc} + {\bar{\chi}}^j \Gamma^{abc}
\chi^j \right ) \right \} . \cr && \label{eq:4fermimatm}
\end{eqnarray}

\vskip 0.1in It is helpful to consider the variation of the matrix
Lagrangian under an infinitesimal supersymmetry transformation,
parameterized by an infinitesimal spinor, $\eta$. As explained
above, we begin by setting: $\eta$ $=$ $\eta(0)$ $+$ $\partial_a
\eta(0) \delta \xi^a$, where we choose $\eta$ itself to transform
as a $U(N)$ singlet. The supersymmetry transformations take the
form \cite{br}:
\begin{eqnarray}
\delta_\eta \Phi &&= {\bar{\eta}} \lambda \cr \delta_\eta
E^a_{\mu} &&=  \half {\bar{\eta}} \Gamma^a \psi_{\mu} \cr
\delta_\eta A^i_{\mu}  &&= \half {\bar{\eta}} \Gamma_{\mu} \chi^i
\cr \delta_\eta B_{ab} &&= \half {\bar{\eta}} \Gamma_{[a}
\psi_{b]} - {{1}\over{g^2}} A^i_{[a} \partial_0 A^i_{b]} \cr
\delta_\eta \psi_{\mu} &&= (\partial_{\mu} - \half \Omega_{\mu
+}^{ab} \Gamma_{ab} )\eta + \half \left ( {\bar{\eta}} \psi_{\mu}
- {\bar{\psi}}_{\mu} \eta \right ) \lambda
 - \half  ( {\bar{\psi}}_{\mu} \Gamma^a \eta ) \Gamma_a \lambda
+ {{1}\over{g^2}} ({\bar{\chi}} \Gamma^{abc} \chi ) \Gamma_{abc}
\Gamma_{\mu} \eta \cr \delta_\eta \chi^i &&= - \half ( \Gamma^{ab}
F^i_{ab} ) \eta + \half \left ( {\bar{\eta}} \chi^i -
{\bar{\chi}}^i \eta \right ) \lambda - \half ({\bar{\chi}}^i
\Gamma^a \eta ) \Gamma_a \lambda \cr \delta_\eta \lambda &&= -
\quart (\Gamma^a D_a \Phi ) \eta + \left ( H_{abc} -
{\bar{\lambda}} \Gamma_{abc} \lambda + {{1}\over{g^2}}
{\bar{\chi}}^i \Gamma_{abc} \chi^i \right )
 \Gamma^{abc} \eta
\quad , \label{eq:susyt}
\end{eqnarray}
where the combination, $\Omega_{\mu +}^{ab}$ $\equiv$
$\omega_{\mu}^{ab}
$$+$${\hat{H}}_{\mu ab}$, and the supercovariant field strength
is defined in Eq.\ (\ref{eq:fieldss}).

\section{The Theory with Sixteen Supercharges}

\vskip 0.1in In preceeding sections, we have described a detailed
prescription for the spacetime reduction of a locally
supersymmetric theory with large $N$ flavor group to a single
point in spacetime, such that the resulting zero-dimensional large
$N$ matrix model Lagrangian manifests all of the local symmetries
of the original continuum field theory. Our prescription is a
modification of Eguchi and Kawai's well-known planar reduction
procedure, which takes into account the necessity for an auxiliary
local tangent space in a covariant Lagrangian formulation of a
gravitational theory describing spinors in a generic curved
spacetime background. As our prototype example, we have analyzed
the case of the heterotic 10d $N$$=$$1$ supergravity-Yang-Mills
Lagrangian with an anomaly-free Yang-Mills gauge group,
$E_8$$\times$$E_8$ or $SO(32)$, of rank $16$. In part, the reason
for this is that a detailed analysis of the low energy spacetime
effective action, up to quartic order in the inverse string
tension, $\alpha^{\prime}$, and in the inverse Yang-Mills coupling
as required by closure under supersymmetry, exists for the
heterotic string supergravity \cite{br}. This comprehensive
analysis is due to Bergshoeff and Roo \cite{br2,br}, building on
the earlier works of \cite{brwn,ch}. The resulting Lagrangian has
been presented in manifestly supersymmetric form, and in terms of
component fields. The equivalence of the dual two-form and
six-form formulations, at least up to quartic order in
$\alpha^{\prime}$, has also been established by these authors.
Partial comparisons have been made, and are in agreement with,
terms in the effective action inferred from direct string
amplitude calculations up to one-loop order \cite{gsloan}.

\vskip 0.1in Following the c.1995 developments in string duality,
we have an enhanced appreciation of the rich structure of the
vacuum landscape of theories with sixteen supercharges. Toroidal
compactification of the 10d heterotic string preserves all of its
supersymmetries, yielding a rich class of theories with sixteen
supercharges, and anomaly-free rank $16$$+$$n$ Yang-Mills gauge
group, in $10$$-$$n$ spacetime dimensions \cite{nw}. The discovery
of the CHL moduli spaces \cite{chl} clarified that the vacuum
landscape is {\em not} simply-connected: these models are
supersymmetry preserving orbifolds of the standard toroidal
compactifications of the heterotic string \cite{cp}. Thus, for
example, in nine spacetime dimensions the vacuum structure of the
theory with sixteen supercharges is already multiply connected: in
addition to the connected vacuum landscape with 17 abelian
one-forms at generic points in the moduli space, we have an
isolated island universe with 17-8=9 abelian one-forms at generic
points.\footnote{Since the orbifold twist becomes trivial in the
noncompact decompactification limit where the requisite massless
gauge bosons are simultaneously recovered, the CHL orbifold is
not, strictly speaking, a disconnected component of the theory
with 16 supercharges \cite{cp}. But we should emphasize that, at
weak coupling, and in the moduli space approximation, each moduli
space describes low-energy physics in a different island universe.
While nonperturbative dynamics can often be invoked to infer the
possibility of tunneling to a different moduli space with {\em
fewer} supersymmetries \cite{others}, no such examples are known
in theories with 32 or 16 supercharges. Note that the mechanism
proposed for a partial breaking of supersymmetry described in
\cite{frey}, giving a theory with 12 supercharges, requires
assumptions about the nature of the theory in the
decompactification limit.} This theory was first identified as an
asymmetric orbifold of the circle compactification of the
$E_8$$\times$$E_8$ heterotic string theory by Chaudhuri and
Polchinski \cite{cp}: the ${\rm Z}_2$ orbifold action is a
supersymmetry-preserving shift in the one-dimensional momentum
lattice, accompanied by the outer automorphism exchanging the two
$E_8$ lattices. The gauge symmetry at generic points in the moduli
space is rank $9$.

\vskip 0.1in We emphasize that there is no known spacetime
dynamics, field-theoretic or string-theoretic, that can repair the
disconnectedness of the moduli space with sixteen supercharges.
Recall that there is no Higgs mechanism in theories with sixteen
supercharges. Thus, while the precise enhanced gauge group can
vary from point to point, the rank of the abelian subgroup is {\em
fixed} for all points in a connected component of the moduli space
\cite{chl}. More precisely, as is clarified by the orbifold
construction \cite{chl,cp,cl}, each isolated component of the
moduli space is characterized by a distinct target-space duality
group entering into specification of the global symmetry algebra
of that island universe. An alternative viewpoint is to realize
that each island universe is an example of a flux compactification
\cite{ps}: one, or more, of the supergravity pform fluxes is
nontrivial, an invariant on a connected component of the moduli
space. The type IIA string duals of the heterotic CHL models with
constant Ramond-Ramond background one-form potential constructed by Chaudhuri
and Lowe \cite{cl} were the earliest known examples of 
compactifications of the type II string theory with nontrivial RR sector,
leading to the generic class of RR/NS flux compactifications. 
While the notion
of isolated universes can be disconcerting, raising the spectre of
the anthropic principle, and banishing hopes of a unique vacuum
state for String/M theory picked by dynamics alone, we have argued
elsewhere that the problem could be one of misinterpretation.

\vskip 0.1in Let us move on to a different aspect of the vacuum
landscape of theories with sixteen supercharges, namely, the fact
that the six different string theories: type I, type I$^{\prime}$,
type IIA, type IIB, heterotic $E_8$$\times$$E_8$, and heterotic
$SO(32)$, each describe a different weakly-coupled limit of the
{\em same} moduli space. Consider the circle compactifications of
all six string supergravities and, for convenience, let us
restrict ourselves to discussion of the standard component of the
moduli space characterized by a rank $16$ anomaly-free Yang-Mills
gauge group. As is well-known, the Lagrangian we have described
above can be mapped by a strong-weak coupling duality
transformation, and suitable field identifications, into that of
the type IB string theory \cite{polwit}. Thus, the $SO(32)$ type I
string theory is the strong-coupling dual of the heterotic string
theory with identical gauge group.

\vskip 0.1in In nine dimensions, and below, the $SO(32)$ and
$E_8$$\times$$E_8$ heterotic string theories are related by a
target space duality:
$R_{9}$$\leftrightarrow$$\alpha^{\prime}/R_9$. What is the type I
strong-coupling dual of heterotic vacua with states in the spinor
representations of the orthogonal groups, as required by the
appearance of exceptional Lie algebras? Fortunately, upon
compactification to nine dimensions, the type I theory can acquire
nonabelian gauge symmetries of nonperturbative origin. Under the
T$_9$ duality, the type I string with its 32 D9branes is mapped to
a type I$^{\prime}$ vacuum with 32 D8branes: additional massless
gauge bosons can arise from the zero length limit of D0-D8brane
strings. Such D0-D8brane backgrounds preserve all sixteen
supersymmetries. The incorporation of D0-D8backgrounds, in
addition to those with only 32 D8branes, permits identification of
type I-I$^{\prime}$ strong-weak coupling duals for {\em all} of
the nine-dimensional ground states of the heterotic string
theories. In particular, this includes compactifications on a
circle of both the $Spin(32)/{\rm Z}_2$ and the $E_8$$\times$$E_8$
heterotic string theories \cite{polwit,bachas,flux}. the inclusion
of the D0-D8brane backgrounds also enables the identification of
the type I-I$^{\prime}$ strong coupling duals for all of the
heterotic CHL moduli spaces with sixteen supersymmetries
\cite{chl,cp,cl}. Furthermore, since type I$^{\prime}$ theory
compactified on $S^1$ is the same thing as M theory compactified
on $S^1$$\times$$S^1/{\rm Z}_2$, these observations are consistent
with the identification of M theory on $S^1/{\rm Z}_2$ as the
strong coupling limit of the $E_8$$\times$$E_8$ heterotic string
theory \cite{hulltown,hw}. We emphasize that the analysis of D0-D8 
type IB backgrounds given by us in \cite{flux}, elevates the equivalence of 
type IB and heterotic string theories to a {\em stringy} equivalence, not
merely at the level of the low-energy field theory limits. The key point is that
the Yang-Mills gauge sector in either case is isomorphic to the massless modes
of a chiral vertex
operator algebra, namely, a chiral conformal field theory.

\vskip 0.1in Most importantly, it is plausible that the success in unifying the
circle-compactified heterotic and typeI-I$^{\prime}$ theories with
sixteen supercharges can be extended to incorporate the type IIA
and type IIB string theories with generic RR backgrounds. 
There exists a nine-dimensional
Lagrangian formulation of the massive type IIA-IIB supergravities
due to Bergshoeff, de Roo, Green, Papadopoulos, and Townsend
\cite{berg} which incorporates the full spectrum of Dbrane
$p$-form potentials \cite{dbrane}, including Roman's IIA
cosmological constant \cite{romans}. By combining
field-dualizations, as well as $S$ and $T$-duality transformations
on the couplings, this Lagrangian can be mapped to {\em any} of
six supergravity theories: the circle-compactified type I, type
IIA, or heterotic string supergravities, the Scherk-Schwarz
reduction of the type IIB string supergravity, or the
$S^1$$\times$$S^1/{\rm Z}_2$ compactification of
eleven-dimensional supergravity. This covers all six vertices of a
modified star diagram linking theories with sixteen supercharges
\cite{hulltown,polbook,mat1}.

\vskip 0.1in With our new prescription for spacetime reduction, we
have shown that the local symmetries of a given field theory
Lagrangian can be preserved in the reduced matrix model. Thus,
there is a precise analog for each field redefinition, or
dualization, of the continuum Lagrangian in the matrix model: the
matrix Lagrangian is only unique upto appropriate dualizations
defined on the matrix variables \cite{mat1}. On the one hand, this
is a beautiful illustration of the underlying 
unity of the different ten-dimensional
string supergravities 
with eleven-dimensional supergravity. But it points to
the importance of understanding the global symmetry algebra: the
identification of a specific, hidden symmetry algebra is what
gives precise meaning to one, or other, class of supergravity/M
theory toroidal compactifications. This observation has been
reiterated recently by West \cite{w2,w1,w4,w5}. But it is not new
to string theory, nor to supergravity: target-space duality
groups, and their conjectured extension to U-dualities
\cite{hulltown}, have been the bulwark of our understanding of
both string and supergravity compactifications. In section 5 of this paper,
we exphasize that the notion of the global symmetry algebra also
provides a precise generalization incorporating {\em all} of the
ground states of String/M theory.
Algebraic considerations were the chief
input into discovery of the disconnected \lq\lq island universes"
with sixteen supercharges by us in \cite{chl}. Each of the CHL moduli 
spaces corresponds to a
supersymmetry 
preserving orbifold of the toroidally compactified
ten-dimensional heterotic 
superstrings \cite{cp}, characterized
by a Lorentzian self-dual lattice with definite isometries.

\vskip 0.1in We began this section by pointing out that, at the
current time, we do not have a comprehensive analysis of the full
covariant Lagrangian--- including all of the fermionic
contributions necessitated by supersymmetry, for any of the low
energy effective Lagrangians other than that of the heterotic
string theory \cite{br2,br}.  For
theories with sixteen supercharges,
this is, of course, the supergravity 
Lagrangian of fundamental significance.

\section{Global Symmetry Algebra of Type IB Supergravity}

\vskip 0.1in We have alluded earlier to the existence of a hidden
symmetry algebra in the matrix Lagrangian that is larger than the
obvious $U(N)$$\times$$G$. In part, there is an $SL(10,{\bf R})$
symmetry, which is the manifest remnant under spacetime reduction
to a single point in spacetime of the Lorentz symmetry group of
the 10d continuum field theory Lagrangian. In the Introduction, we
have already explained the simple rationale for expecting the
symmetry algebra of an 11d M theory with 32 supercharges
to be $E_{11}$ $=$ $E_8^{(3)}$, the
rank eleven algebra known as the very-extension of the finite
dimensional Lie algebra $E_8$ \cite{w4}.\footnote{A pedagogical
review of some recent directions in the study of the symmetry
algebra of theories with 32 supercharges and, in particular, of the very-extended Lie
algebras, is given in \cite{matn6}. The paper addresses recent
results of West and collaborators in this context \cite{w4}.} 

\vskip 0.1in The
{\em nonmaximal} 10d supergravity is a theory with sixteen supercharges.
It also leads to maximal supergravity-Yang-Mills theories in ten
dimensions, the low energy limits of the anomaly-free
heterotic and type I superstring
theories \cite{chl}.
In the notation of \cite{w4,matn6}, we have the following symmetry
generators:
\begin{equation}
K^a_b , R , R^{c_1 c_2} , R^{c_1 \cdots c_6} , R^{c_1 \cdots c_8}
 \quad .
\label{eq:typei}
\end{equation}
The heterotic supergravity theory has zero-form dilaton and NS
two-form potentials, plus their ten-dimensional Hodge duals,
respectively, six-form and eight-form supergravity potentials. The
$K^a_b$ are the generators of $GL(10,{\bf R})$ in the notation of
\cite{w4}. The commutator algebra of these generators was given in
\cite{w5}. Not surprisingly, we will find that it agrees precisely
with the algebra that can be inferred from an appropriate chirality 
projection on the global symmetry algebra of the type IIB
supergravity. This reflects the well-known connection between
these two superstring theories, following the orientation projection 
to the symmetric combination
 left-moving and right-moving modes on the worldsheet of the
 type IIB superstring 
\cite{polbook}.

\vskip 0.1in The global symmetry algebra of the type IB
supergravity can therefore be identified by performing a chiral projection
on the global symmetry algebra, ${\cal G}_{IIB}$, of the 10d type
IIB supergravity, which was also obtained in a recent work of
Schnakenburg and West \cite{w2}. As shown in \cite{matn6}, by
setting the extra forms to zero in Eqs.\ (1.1-.3) of \cite{w2}, we
find the usual $GL(10,{\bf R})$ algebra, extended by translations:
\begin{equation}
[K^a_b , K^c_d ] = \delta_b^c K^a_d
- \delta_d^a K^c_b , \quad\quad [K^a_b , P_c ] = \delta_c^a P_b ,
\quad\quad [K^a_b , R^{c_1 \cdots c_p } ] = \delta_b^{c_1} R^{ac_2
\cdots c_p} + \cdots  \quad , \label{eq:gl10}
\end{equation}
plus the simplified algebra of 0, 2, 6, and 8-form generators:
\begin{equation}
[R , R^{c_1 \cdots c_p} ] = d_p R^{c_1 \cdots c_p}  , \quad
[R^{c_1 \cdots c_p} , R^{c_1 \cdots c_q} ] = c_{p,q} R^{c_1 \cdots
c_{p+q}}
 \quad . \label{eq:extrabI}
\end{equation}
Comparing with the IIB result given in Eq.\ (1.3) of \cite{w2},
the remnant non-vanishing structure constants take the simple
form:
\begin{equation}
d_{q+1} = - \quart (q-3) ,  ~ q=1,5, \quad\quad \quad c_{2,6} =
\half \quad . \label{eq:strbI}
\end{equation}
The algebra we obtain is in precise agreement with that of the 10d
$N$$=$$1$ heterotic supergravity given in Eq.\ (1.4) of \cite{w5}.
Let us denote this algebra as ${\cal G}_{IB}$. We emphasize that,
thus far, we have not included the Yang-Mills gauge sector of the
nonmaximal 10d supergravity-Yang-Mills theory.

\vskip 0.1in Let us now address the question of how the global
symmetry algebra of the heterotic-type I nonmaximal supergravity
relates to the symmetry algebra of M theory with 32 supercharges. West has provided
mounting evidence in favor of the conjecture that the symmetry
algebra of M theory is the rank eleven very-extended algebra
$E_{11}$ \cite{w4}. A review, and our own assessment, of West's
arguments appears in \cite{matn6}. If we compare the generators
and commutation rules given above with the Chevalley basis for the
algebra $E_8^{(3)}$, written in either its IIA or IIB guise as
shown in \cite{w4}, we find that we are missing some of the
positive root generators in either formulation. We have all of the
generators, $E_a$$=$$K^a_{a+1}$, $a$$=$$1$, $\cdots$, $9$, of
$SL(10,{\bf R})$. In the IIA formulation, given in Eq.\ (4.4) of
\cite{w4}, we are missing the roots corresponding to the R-R
one-form, and NS-NS twoform, namely, $E_{10}$$=$$R^{10}$, and
$E_{11}$$=$$R^{910}$. In the IIB formulation, we are missing the
roots labelled $E_{9}$$=$$R_1^{910}$, and $E_{10}$$=$$R_2$,
arising, respectively, from the NS-NS two-form potential, and R-R
scalar. It is clear we cannot build a full $E_{8}^{(3)}$ algebra
from the restricted set of generators in ${\cal G}_{IB}$.

\vskip 0.1in In \cite{w5}, it was pointed out that a different
rank eleven very-extended algebra, namely, the very-extension of
the $D_8$ subalgebra of $E_8$, can be spanned by the generators of
${\cal G}_{IB}$. We should note that such a construction is
somewhat unmotivated from the viewpoint of any relationship to the
type II theories, to eleven-dimensional supergravity, or to M
theory: the authors of \cite{w5} make the choice
$E_a$$=$$K^a_{a+1}$, $a$$=$$1$, $\cdots$, $9$, $E_{10}
$$=$$R^{910}$, and $E_{11}$$=$$R^{5678910}$. This choice of simple roots
is shown to generate the very-extended algebra $D_8^{(3)}$.
Appending a one-form generator to this set converts the
$D_8^{(3)}$ algebra to $B_8^{(3)}$ \cite{w5}. However, it should
be noted that, since the two-form and six-form potentials are
Hodge dual to each other in ten dimensions, a construction which
includes both in the simple root basis is quite different in
spirit from that of the $E_{8}^{(3)}$ algebras underlying the IIA
and IIB theories \cite{w4}. On the other hand, the generators of
 very-extension
of $D_8$ appears to encapsulate the basic physics of 
the supergravity sector of the 
renormalizable superstring theories remarkably well: 
gravity, the dilaton and antisymmetric two-form potential, plus their
10D electric-magnetic Poincare-Hodge duals.

\vskip 0.1in One possible direction towards extending the
$D_8^{(3)}$ to a full $E_8^{(3)}$ algebraic structure in supergravity
theories with sixteen supercharges is to consider compactifications. 
Since we already have the
requisite two-form potential, respectively, the R-R, or NS-NS,
two-form of the type IB, or heterotic, supergravities, our goal
will be to identify a one-form potential that can play the role of
the positive root generator labelled $E_{10}$ in the IIA
formulation of the $E_8^{(3)}$ algebra. A hint is provided by our
understanding of the Duality Web linking the zero slope limits of
the circle compactifications of the type I, type IIA, IIB, and
heterotic string theories, along with M theory compactified on
$S^1$$\times$$S^1/{\rm Z}_2$. Upon compactification on a circle,
the heterotic string theories acquire an abelian one-form
potential, namely, a Kaluza-Klein gauge boson. This perturbative
gauge field is known to play a crucial role in six-dimensional
weak-strong IIA-heterotic string-string duality: it maps under a
weak-strong coupling duality to the R-R one-form potential of the
IIA string theory compactified on K3.\footnote{To be more precise,
compactifications of the IIA theory on K3 are described by a
$(19,3)$ cohomology lattice characterizing classical K3 surfaces.
It is the quantum extension to a $(20,4)$ quantum cohomology
lattice, as a consequence of introducing a flux for the R-R
one-form potential, that completes the isomorphism of the IIA
theory compactified on K3 to the heterotic string compactified on
$T^4$. The quantum cohomology lattice is identified with the
$(20,4)$ Lorentzian momentum lattice of the heterotic string
\cite{sen1,harvs,cl}. The heterotic theory has 20 abelian
one-forms. One of these is distinguished as the partner of the R-R
one-form of the IIA theory, and this is the Kaluza-Klein gauge
field we have in mind.} As the simple root generator labelled
$E_{11}$ in the IIA formulation of $E_8^{(3)}$, we might choose,
respectively, the NS-NS two-form potential of the IIA string or
the two-form potential of the heterotic string. Note that these
are mapped to each other under string-string duality. As an aside,
the reader may wonder why we had to compactify all the way to six
dimensions to see these equivalences, but this methodology is in
keeping with how the Cremmer-Julia hidden symmetries of the type
II theories were discovered. The full global symmetries only
become manifest in the dimensionally-reduced supergravity
theories, but this can be taken as a hint towards discovering a
higher-dimensional correspondence. In summary, identifying the
Kaluza-Klein one-form, and the two-form potential, as two of the
missing positive root generators in the Chevalley basis might
plausibly allow one to demonstrate an $E_8^{(3)}$ global symmetry
algebra in the circle compactifications of the heterotic
supergravities.

\vskip 0.1in It is important to notice that the
supergravity structure of the CHL orbifolds is identical with that
of the circle compactifications. Indeed, if we are correct in our
expectation that the circle-compactified theory has an underlying
$E_8^{(3)}$ global symmetry algebra, then this would also be true
of the CHL orbifold. The distinction between these two theories lies
only in the Yang-Mills sector: they differ in the rank of the gauge
group at generic points in the moduli space, respectively, 17, and
9, and we are arguing that the additional Kaluza-Klein gauge
bosons contribute to the extension of $D_8^{(3)}$ to $E_8^{(3)}$.
Are the precise nonabelian enhanced gauge symmetry groups
relevant to this discussion? It has become customary to think not,
since it is well-known that the precise nonabelian enhancement
varies from point to point in the moduli space. Conventionally,
this multiplicity of enhanced symmetry points can be inferred by
analysis of 
the perturbative T-duality groups.
Now it is clear that the full 
T-duality group of the toroidally compactified heterotic string can 
never be accommodated within $E_8^{(3)}$: we emphasize that
this is unlike the smaller T-duality group of the toroidally compactified
type II superstrings. As shown by Lu and Pope \cite{lp}, the Cremmer-Julia
group always contains the T-duality group in this case. The Yang-Mills
sector of the heterotic and type I string theories therefore poses
a puzzle for the Cremmer-Julia conjecture: how should one incorporate
the one-form generators of the Yang-Mills sector into the hidden symmetry
algebra?

\vskip 0.1in In summary,  in light of what we have learned
from the survey of the hidden symmetry algebras of nine,
ten, and eleven, dimensional supergravity theories in \cite{matn1},
and given the
striking evidence that the rank eleven very-extension of the Lie
algebra $E_8$ incorporates the full spectrum of NS-NS and R-R
charges, including the crucial D8brane and D(-2)brane charges, we
will conjecture that the hidden symmetry algebra of supergravity
theories with sixteen supercharges takes the form
$G^{\prime}$$\times$$ G$, where $G$ is the finite-dimensional 
Yang-Mills gauge symmetry, and $G^{\prime}$ {\em could} be as large as $E_8^{(3)}$. We
will conjecture further that nonperturbative String/M theory is the
realization of this algebra
on unitary $N$$\times$$N$ matrices. Further elucidation of our
conjecture is left to
future work.

\section{Conclusions and Future Directions}

\vskip 0.1in A proposal as radical as that described in this paper
\cite{mat1} has few concrete conclusions in comparison with the
Pandora's box of fascinating questions it opens up for future
investigation. The significant achievements in giving a concrete
proposal for an emergent spacetime geometry are already summarized
in the Introduction. Let us focus in this concluding section on
the most accessible of the open questions raised by our work:

\begin{itemize}

\item{It is always helpful to gain what insight may be available
from detailed calculations in toy matrix models. In this context,
it is noteworthy that a refreshed perspective on nonperturbative
effects in string theory has emerged from recent work on
noncritical type 0 string theory in a two-dimensional target space
\cite{mmn}, and on {\em minimal} string theory, namely, the (p,q)
conformal minimal models coupled to Liouville gravity
\cite{minimal}. A new result \cite{kawai1}, motivated in part by
these developments, is the elegant numerical computation of the
nonperturbative Dinstanton contribution to the perturbative
genus-by-genus expansion of the free energy of the
zero-dimensional matrix model, the fully nonperturbative
description of two-dimensional quantum gravity coupled to unitary
matter with $c<1$ \cite{mm,shenk}.\footnote{We note in passing
that it should be possible to carry out a similar analysis to
\cite{kawai1} for the rather interesting zero-dimensional reduced
matrix model describing the coupling to $c=1$ matter. This exactly
solvable matrix model is known as the Penner model \cite{distv},
and the exact solution to the Lagrangian matrix path integral
obtained by us using the orthogonal polynomial method appears in
\cite{cl2}.} An important general implication of the analysis in
\cite{kawai1} relevant for critical string/M theory, is that {\em
the matrix path integral formalism preserves the normalization of
the Dinstanton action}: this is nonperturbative data, above and
beyond the form of the action inferred from
the asymptotic large order behavior of the solution to the matrix model loop
equations \cite{david}.

\vskip 0.1in This observation has a striking parallel in the
worldsheet formalism of perturbative string theory: the
Weyl-invariant path integral formalism for open and closed string
amplitudes preserves the {\em normalizations} of string amplitudes
\cite{poltorus,fuk,zeta}, a powerful framework that goes beyond
what can be extracted from the operator formalism of conformal
field theory, and of light-cone gauge string theory. Thus, both
because of its powerful computational elegance, demonstrated by
analyses such as \cite{kawai1}, and because it enables addressing
issues in quantum gravity, and quantum cosmology, that can only be
sensibly formulated in an off-shell, or semi-classical, framework
eschewing the use of S-matrices, the development of mathematical
techniques that make the Lagrangian matrix path integral directly
amenable to analysis needs to be a central goal of future work in
matrix model techniques. The strong evidence for a hidden symmetry
algebra in the spacetime reduced matrix Lagrangians of relevance
to M theory suggests that we need to develop a completely new
strategy towards their analytic solution, and which is likely to
be rooted in this algebraic structure. The necessary mathematical
techniques could turn out quite different from methods that have
proven effective in the toy matrix models.}

\item{Having motivated the necessity to focus attention on the
hidden symmetry algebra of the theory with sixteen supercharges,
let us highlight the gaps in our current understanding that remain
to be filled. The identification of the global
symmetry algebra of the type
I-I$^{\prime}$-heterotic string theories with correct incorporation 
of the Yang-Mills sector 
needs to be completed. In particular, West has made the
interesting observation that the IIA-IIB T-duality transformation
simply reflects the bifurcation symmetry of the $E_8^{(3)}$ Dynkin
diagram at its central node, interchanging the Dynkin diagrams of
its two inequivalent $A_{9}$ subalgebras \cite{w4}. This argument
can clearly be adapted to the T-duality symmetry relating the type
I and type I' string theories. Or, to that relating the two
circle-compactified heterotic string theories. The details of our
broad conjectures need to
be verified.}

\item{ It should be noted that our
new perspective on the hidden symmetry algebra takes seriously the
strong-weak dualities linking the heterotic, type IB, and type
IIA, string theories: a theory of sixteen supercharges is
self-dual, and it would be meaningless to have different hidden
symmetry algebras pertaining to the different string theories.
Thus, while the perturbative T-duality group of the type II theory
is, in fact, incorporated in $E_{11}$, the Lorentzian extension 
familiar from the toroidally-compactified
heterotic string {\em cannot} be contained within $E_{11}$. Our
conjecture is that the hidden symmetry algebra of the
type I-heterotic supergravity with sixteen supercharges, $G^{\prime}$,
will be
unchanged for {\em all} of the CHL theories. Of course, the
abelian subgroup of the nonabelian gauge symmetry characterizing
generic points in a given moduli space, and hence $G$, will be
different for each of the latter CHL theories. It just so happens
that in nine dimensions there are no Wilson lines that permit
either a breaking, or enhancement, of the $E_8$ Yang-Mills gauge
symmetry.\footnote{I thank Arjan Keurentjes for requesting this
clarification.}}

\item{Once the precise nature of the hidden symmetry algebra of
the type I-I$^{\prime}$-heterotic theories with sixteen
supercharges has been pinned down, and which we have conjectured
will take the form $G^{\prime}$$\times$$G$, what new physics
becomes accessible? It is a remarkable fact that there exists a
{\em unique} assignment of phases in the bosonic $E_8^{(3)}$
algebra corresponding to an eleven-dimensional theory with,
respectively, Minkowskian (1,10), or Euclidean (0,11) signature
\cite{hulld,keurentjes}. As has been shown by Keurentjes
\cite{keurentjes}, every other self-consistent choice of phases
for $E_{11}$ results in a spacetime with two, or more, timelike
directions. Furthermore, the Euclidean case also corresponds to a
bosonic $E_{11}$ algebra with natural physical interpretation as
the symmetry algebra of M theory at finite temperature. This
Euclidean symmetry algebra might be of great interest in the
context of M theory cosmology, as we now explain.

\vskip 0.1in We will offer a suggestive interpretation for the
principle subalgebra of the Euclidean signature bosonic $E_{11}$
algebra. As expected from generic considerations \cite{nico},
every Lorentzian Kac-Moody algebra has a principal $SO(1,2)$
subalgebra, and it is natural to seek its physical interpretation.
Based on our understanding of the String/M duality web in 11, 10,
and 9, spacetime dimensions, and given the pivotal role played by
the nine-form potential and its (-1)-form dual, it is natural to
identify the parameters of the $SO(1,2)$ subalgebra, roughly, as
follows.\footnote{We should emphasize that this is very rough
intuition. Thus, we are not making any clear statement on the
precise topology of the group, discrete identifications might be
necessitated.} Labelling them as $R_0$, $R_{10}$, and $R_9$,
respectively, suggests a natural identification with inverse
temperature, $\beta$, string coupling, $g$, and cosmological
constant $M$. The latter is Roman's mass parameter, later
interpreted by Polchinski as D8brane charge \cite{romans,dbrane}.
Notice that the two-parameter subspace $(\beta, g)$, whose rough
correspondence with the radius of coordinates $(X^0,X^{10})$ is
well-known, is also the classic phase space parameterization
relevant for the study of the dynamics of a finite temperature
gauge theory \cite{decon}. Supplementing this with the
cosmological constant gives a natural three-parameter phase space
relevant for discussions of M theory cosmology: the thermal
dynamics of the Universe, inclusive of gravity \cite{decon}. It
should be emphasized that the principal $SO(1,2)$ algebra should
{\em not} be confused with the corresponding subalgebra of the
spacetime Lorentz algebra.

\vskip 0.1in In recent work on string thermodynamics \cite{decon},
we have pointed out that there is a fundamental conceptual barrier
to proposals for a {\em microcanonical} description of the string
ensemble: perturbative string theory is inherently a
background-dependent theory. Thus, we cannot escape the \lq\lq
heat-bath" represented by the spacetime geometry, and additional
background fields: {\em any} self-consistent discussion of string
thermodynamics must therefore be relegated to the canonical
ensemble. Fortunately, there is a first-principles framework for
the canonical ensemble provided by the world-sheet path integral
formalism, originally pointed out in \cite{poltorus}. On the other
hand, from the perspective of quantum cosmology, the ensemble of
interest {\em is} the microcanonical ensemble of the fundamental
degrees of freedom: the Universe is a closed system, and there is
no \lq\lq heat bath" one can point to. The reduced matrix models
we have described in this paper offer a self-consistent starting
point in which to formulate the microcanonical ensemble of the
fundamental degrees of freedom. This opens up the exciting
possibility of a genuinely nonperturbative formulation for black
hole thermodynamics and quantum cosmology.}

\item{The detailed understanding of the {\em fermionic} sector of
the ten and eleven dimensional supergravities with 32, and 16,
supercharges in terms of the hidden symmetry algebra, $E_8^{(3)}$,
or $G^{\prime}$$\in$$E^{(3)}_8$, respectively, needs to be
completed. Preliminary work constructing the standard 
double-sided, spinorial
representations of the SO(10,1), or SO(9,1), Lorentz algebras from 
the appropriate representation of the Cartan sub-algebra of
the Lorentzian Kac-Moody algebra is under way
\cite{ns,kb}.\footnote{We should note here the
recent paper \cite{nk} which focusses on the rank 10 hyperbolic
Kac-Moody algebra $E_{10}$: $E_{10}$ has an $A_9$ subalgebra, but
not the full $A_{10}$ expected in a theory that incorporates a
background with 11d supergravity, and the generalized
electric-magnetic duality described in \cite{hodge}. Not
surprisingly, the analysis of \cite{nk} cannot incorporate a
space-filling D9brane since there is no corresponding rank ten
generator, nor its magnetic dual, and $E_{10}$ does not therefore appear to be a
viable candidate for the hidden symmetry algebra of string/M
theory. I thank H.\ Nicolai for clarifications.}}

\item{As an aside, we should note that an important issue raised
both by our focus on the principal three-parameter subgroup of
$E_{8}^{(3)}$, and by its realization in a locally supersymmetric
unitary matrix model, is the possibility of an undiscovered
relation to the famous {\em supermembrane theory}, a conjectured
theory of fundamental supermembranes \cite{sm,dhopn}. Does the
locally supersymmetric matrix model represent a regularization of
the three parameter manifold of the principal subgroup, analogous
to the regularization of the worldvolume of the supermembrane
provided by the rigid unitary matrix model \cite{dhopn}? These are
open questions that might shed light on the large N continuum
limit of the reduced matrix model.}

\item{Finally, coming to the crucial open questions in the matrix
model framework, there is the issue of what comes beyond leading
order in large $N$ in the matrix model: what is the significance
of the off-diagonal elements of the variables in the reduced
matrix model? Notice that there is an obvious extension to the
notion of the double-scaling limit familiar from the $c=1$ matrix
model: namely, $\lim_{N \to \infty , g \to 0}$, with $g^{\alpha}
N^{\beta}$ held fixed. The parameters $(\alpha, \beta)$ take an
appropriate range of values for members in the discrete series of
the gravitationally-dressed unitary conformal field theories with
central charge $c$$\le$$1$ \cite{mm}. The generalization to large
$N$ limits with multiple-scaling was pointed out in our earlier
works \cite{mat1}. Since we have a full range of background
fields, $(g=e^{\bar{\phi}}, {\bar{A}}_{c_1}, {\bar{C}}_{c_1},
{\bar{C}}_{c_1c_2}, \cdots , {\bar{C}}_{c_1 \cdots c_9})$, where $
g$$=$$(M_{11}R_{10})^{3/2}$, and the single mass scale, $M_{11}$,
 there are
many possible inequivalent, multiple-scaling limits: a suitable
combination of powers of $N$, $M_{11}$, and the background fields,
can be held fixed, in the limit that we take $N$ $\to$ $\infty$.
Here, $M_{11}$ has been taken to be the eleven-dimensional Planck
mass. The precise powers of $M_{11}$ that enter into taking the
large $N$ limit can vary, depending on whether we wish to match to
an eleven, or ten-dimensional, continuum field theory. For
example, the ten-dimensional string mass scale is related as
follows: $m_s$ $=$ $\alpha^{\prime -1/2}$ $=$ $M_{11}^{3/2}
R_{10}^{1/2}$. We should emphasize the fact that it was essential
that the matrix Lagrangian framework allow for a wide range of
inequivalent large $N$ limits, since it would not otherwise be
possible to explain the multitude of known effective dualities
relating M theory ground states.}

\item{Perhaps the most important open question is the comparison
of corrections to the large $N$ limit of the matrix model,
calculated with the specific choice of scaling appropriate for
matching to a particular string supergravity, with the higher
order in $\alpha^{\prime}$ corrections to the string spacetime
effective Lagrangian. We should remind the reader that the precise
form of the low energy spacetime effective Lagrangian for string
theory has not been systematically calculated beyond quartic order
in the inverse string tension \cite{br2,br}, and that too only in
the case of the heterotic string. This is unfortunate, given that
the techniques for the systematic derivation of these terms from
string amplitude calculations, or based on duality symmetries of
the effective action, have been known for many years. In the past,
this was explained by the absence of any direct physical
motivation for a {\em comprehensive} analysis. For example, it was
common to focus on the particular subset of terms that had the
potential to mediate some new physics beyond the standard model.
But given our current understanding of String/M theory, it would now seem
that there is {\em strong motivation} for a renewed effort at
obtaining a comprehensive analysis of the spacetime effective
Lagrangian. We emphasize that it is only at higher orders in the
$\alpha^{\prime}$ expansion that we can successfully test any
conjectured nonperturbative proposal for String/M theory beyond agreement
with the supergravity prediction.}

\end{itemize}

\noindent In summary, we believe this could be the beginning of an
exciting period in the search for a more fundamental description
of String/M theory that transcends its weakly-coupled perturbative
limits. 

\vskip 0.2in \noindent{\bf ACKNOWLEDGMENTS}

\vskip 0.1in I am grateful to Hikaru Kawai for the opportunity to
make a series of stimulating visits to Kyoto (Japan) in the Fall
of 2001, where the work reported in \cite{mat1} was initiated, and
the Kavli Inst of Theoretical Physics, where \cite{mat1} was
completed. I also thank E.\ Cremmer, V.\ Kazakov, and I.\ Kostov,
for arranging the visit to the Paris area institutions in the
Summer of 2002, where this work continued. The derivation of the
matrix Lagrangian given in this paper was formulated in Spring
2004. I would like to thank Thibault Damour, at the IHES, for an
important discussion. Bernard de Witt and Hermann Nicolai also
contributed greatly to my understanding; I thank them for the
hospitality extended to me at the Albert Einstein Institut,
Potsdam. I would like to thank Axel Kleinschmidt for questions and
correspondence, and Emil Martinec, Savdeep Sethi, and Cosmas
Zachos for helpful discussions and interest. Finally, I would like to
thank Greg Moore, Peter West, Jim Lepowsky, Lisa Carbone, 
Hisham Sati, Elizabeth Jurische, and other participants of the 2005
Rutgers Math Workshop on Groups and Algebras in M Theory, for 
deepening my understanding of Lorentzian Kac-Moody algebras.

\end{document}